\documentclass[prb,a4paper,longbibliography,twocolumn,superscriptaddress,english,showpacs]{revtex4-1} 
\usepackage[english]{babel}
\usepackage{xcolor} 
\usepackage{graphicx} 
\usepackage{amsmath} 
\usepackage{amssymb}
\usepackage[normalem]{ulem}
\usepackage[textsize=tiny]{todonotes}

\definecolor{darkblue}{rgb}{0.1,0.2,0.6} \definecolor{darkred}{rgb}{0.8,0.1,0.2}
\usepackage[colorlinks,citecolor=darkblue,linkcolor=darkred,urlcolor=darkblue]{hyperref}
\usepackage[all]{hypcap} 

\makeatletter \adddialect\l@English\l@english \makeatother

\newcommand{\bra}[1]{\langle\,#1\,|} 
\newcommand{\ket}[1]{|#1\rangle}

\newcommand{\mat}[1]{\boldsymbol{#1}}
\renewcommand{\vec}[1]{\boldsymbol{#1}}

\newcommand{\E}{\mathrm{e}} 
\newcommand{\D}{\mathrm{d}}

\newcommand{\cf}{\textit{cf.} } 
\newcommand{\ie}{\textit{i.e.} } 

\newcommand{\vs}{\textit{vs.} } 

\newcommand{\etc}{\textit{etc.} }

\newcommand{\tr}{\mathrm{Tr}}

\begin{document}
\title{Bimodal entanglement entropy distribution in the many-body localization transition }
\author{Xiongjie Yu}
\affiliation{Institute for Condensed Matter Theory and Department of Physics, University of Illinois at Urbana-Champaign, Urbana, IL 61801, USA}
\author{David J. Luitz}
\affiliation{Institute for Condensed Matter Theory and Department of Physics, University of Illinois at Urbana-Champaign, Urbana, IL 61801, USA}
\author{Bryan K. Clark}
\affiliation{Institute for Condensed Matter Theory and Department of Physics, University of Illinois at Urbana-Champaign, Urbana, IL 61801, USA}
\email{dluitz@illinois.edu}
\date{\today}

\begin{abstract} 
We introduce the cut averaged entanglement entropy in disordered periodic spin chains and prove it
to be a concave function of subsystem size for individual eigenstates.
This allows us to identify the entanglement scaling as a function of subsystem size for individual
states in inhomogeneous systems.
Using this quantity, we probe the critical region between the many-body localized (MBL) and ergodic phases
in finite systems. 
 In the middle of the spectrum, we show evidence for bimodality of the entanglement distribution in the
MBL critical region, finding both volume law and area law eigenstates over disorder realizations as well as
within \emph{single disorder realizations}.   The disorder averaged entanglement entropy 
in this region then scales as a volume law with a coefficient below its thermal value.
We discover in the critical region, as we approach the thermodynamic limit,
that the cut averaged entanglement entropy density falls on a one-parameter family of curves.  
Finally, we also show that without averaging over cuts the slope of the entanglement entropy \vs subsystem
size can be negative at intermediate and strong disorder, caused by rare localized regions in the
system.
\end{abstract} 


\maketitle

\section{Introduction}

\begin{figure}[h]
    \centering
	\includegraphics{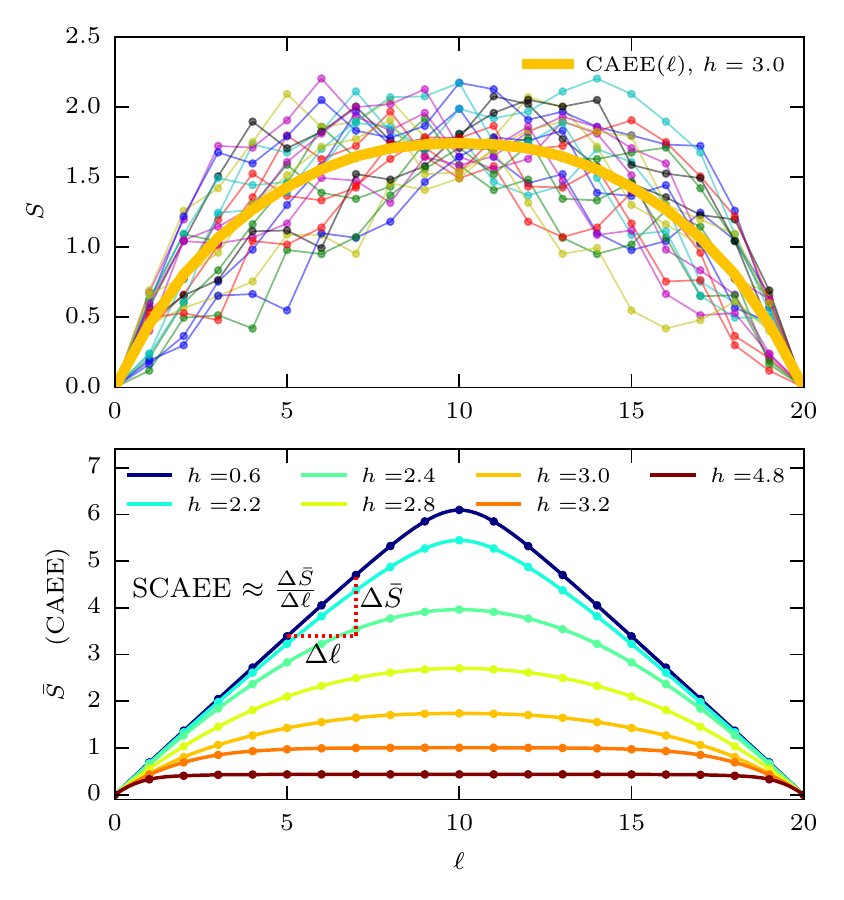}
    \caption{\textbf{Top:} von Neumann entanglement entropies ($S$) for different fixed left cut positions as
        a function of subsystem size $\ell$ of a single eigenstate for one sample with system size
        $L=20$ and disorder strength $h=3.0$. Notice that the $S(\ell)$ curves are in general not
        differentiable. However, the cut-averaged entanglement entropy (CAEE, defined in Eq. \eqref{eq:CAEE}) is a smooth and concave function of $\ell$. 
        \textbf{Bottom:} Typical CAEE (denoted as $\bar{S}(\ell)$) sampled from different disorder
        strengths, for periodic Heisenberg chains of length $L=20$. Each curve is generated from a
        single eigenstate of \emph{one} disorder realization. The slope of CAEE, abbreviated as SCAEE and defined in Eq. \eqref{eq:SCAEE}, can directly probe the volume law or area law scaling behavior of an eigenstate.}
    \label{fig:averageEE}
\end{figure}

The many body localization (MBL) transition is a dynamical phase transition driven by the interplay
of strong interactions and disorder\cite{basko_metalinsulator_2006,gornyi_interacting_2005}. 
While disordered noninteracting systems in one dimension localize for arbitrarily 
small disorder\cite{anderson_absence_1958}, the presence of interactions thermalizes
quantum systems up to a finite critical disorder strength. 
This thermalization is typically expected to occur through the 
eigenstate thermalization hypothesis\cite{deutsch_quantum_1991,srednicki_chaos_1994} (ETH),
which states that local few-body observables become a smooth function of the energy.
Furthermore, their canonical expectation values match those obtained from the mixed state thermal
density matrix at 
inverse temperature $\beta$, chosen such that the thermal expectation value of the energy $\langle H
\rangle_\beta = \frac{1}{Z} \tr\left( \E^{-\beta H} H \right)$ is equal to the eigenenergy of the
state. To satisfy this, 
the reduced density matrix of individual eigenstates at energy $E$ becomes equal to the
thermal density matrix, leading to an extensive (volume law) entanglement entropy.

On the other hand, in the MBL phase at stronger disorder the ETH is no longer valid. 
While ETH is a consequence of quantum chaos (a generic
feature of interacting nonintegrable systems\cite{montambaux_quantum_1993}), the MBL
phase shows features of integrability, most prominently signaled in a change of the spectral
statistics that was explored in several pioneering works
\cite{jacquod_emergence_1997,georgeot_integrability_1998,song_low-energy_2000} and are now a
standard measure for the detection of MBL
\cite{oganesyan_localization_2007,pal_many-body_2010,luitz_many-body_2015,serbyn_spectral_2016,monthus_level_2016}.
The integrability in the MBL phase is due to an emergent extensive number of local 
conserved quantum operators
\cite{serbyn_local_2013,imbrie_many-body_2014}, which prohibit thermalization and lead to a subextensive (area law) entanglement
entropy; this has been numerically verified in many
studies\cite{bauer_area_2013,luitz_many-body_2015,luitz_long_2016,yu_finding_2015,khemani_obtaining_2015,lim_nature_2015}. In particular, after the transition
to the full MBL regime, nearly all eigenstates exhibit area law entanglement\cite{bauer_area_2013,yu_finding_2015,khemani_obtaining_2015,lim_nature_2015}
at arbitrarily high energies. 

While the ETH  is typically only expected\cite{garrison_does_2015} to be valid for subsystems of size $\ell$ such that $\ell/L\rightarrow 0$, evidence for the 
MBL transition typically considers the average half-cut entanglement entropy averaged over the ensemble
of disorder realizations, finding it to either follow a
volume law (in the ergodic region) or an area law (in the MBL region).  In this work, we will consider 
subsystem sizes which are a constant fraction ($<\frac{1}{2}$) of the entire system.

Although the many-body localized and ergodic phases have been heavily studied, the transition between
them is still subject to debate. On general grounds, it has been argued that the scaling of the
entanglement entropy at the critical point should follow a volume law \cite{grover_certain_2014},
whereas phenomenological RG calculations point to a strongly fluctuating behavior
\cite{vosk_theory_2015}. Other studies point to strong multifractal behavior at the critical point
\cite{monthus_many-body-localization_2016} and the possibility of a power law scaling  of the
entanglement entropy as $S\propto L^\alpha$ with an exponent $\alpha<1$ (\ie slower than volume law)
\cite{monthus_many_2016}.

In this work, we consider the properties of the transition (as well as the adjoining phases), by focusing on the 
\emph{cut-averaged entanglement entropy} (CAEE) $\bar{S}(\ell)$ (as in Eq. \eqref{eq:CAEE}) and its slope (SCAEE) (as in 
Eq. \eqref{eq:SCAEE}) of subsystems with size $\ell$ at a fixed ratio of $\ell/L$. Cut averaging here 
literally means averaging over all subsystems of a particular size $\ell$. The CAEE for any eigenstate of a 
periodic system is a concave function of $\ell$ (\cf lower panel of Fig. \ref{fig:averageEE} as
opposed to the upper panel for the entropy without cut average). We prove this using strong subadditivity
(SSA) in Sec. \ref{sec:continuous}. The CAEE and SCAEE can be used to directly identify the volume
or area law scaling in single eigenstates.

Using these concepts, we study the distribution of the SCAEE over disorder realizations. This distribution appears Gaussian at 
weak disorder, while at moderate disorder in the ergodic region the distribution is generically non-Gaussian for 
system sizes we can access, a feature that has been observed also in the distributions of the diagonal 
\cite{luitz_long_2016} and off-diagonal \cite{luitz_anomalous_2016} matrix elements of local operators in the 
eigenbasis of the Hamiltonian. On the MBL side, the distribution of the SCAEE is peaked at zero slope and has 
an exponential tail. In addition, in the MBL and ergodic phases the variance of this distribution
gets smaller as a function of system size $L$, which naturally suggests that the variance approaches
zero in the thermodynamic limit (TDL);  this would leave all eigenstates to follow either an area
law (SCAEE equal to 0) or a volume law (SCAEE close to $\ln(2)$). 

In the transition region, we also find that as the system size grows, $\bar{S}/L$ at a fixed subsystem ratio of 
$\ell/L$ appears to approach a one-parameter family of curves, which can be parameterized by the value of the CAEE or the SCAEE at any $\ell/L$.

Most interestingly, in the critical regime we find that 
the distribution of the SCAEE is bimodal both over multiple disorder realizations as well 
as for \emph{single disorder realizations}. The variance of the SCAEE distribution 
in the transition region seems to grow with system size when considered over disorder realizations.
Our system sizes are too small to pin down its maximal value, but they are
consistent with (among other possibilities) the maximal variance possible which would lead to half the states
 having zero SCAEE and half having maximal SCAEE. 
This scenario would lead to an entanglement entropy at the transition which scales as a volume law with half its thermal value.

\section{Strong Subadditivity and Entanglement Entropy under Periodic Boundary Conditions}
\label{sec:continuous}

Strong subadditivity (SSA) is a theorem of entropy, applicable to both classical and quantum
entropies\footnote{With R\'enyi index 1}. SSA of the von Neumann entropy was proved by E.H. Lieb and M.B. Ruskai in 1973
\cite{lieb_proof_1973}, and can be formulated as many equivalent inequalities\cite{lieb_proof_1973,lieb_convexity_1975,lindblad_completely_1975,ruskai_inequalities_2002}. The von Neumann entanglement entropy $S(A)$ quantifies the entanglement of subsystem $A$
with the rest of the system and is obtained from the reduced density matrix, given by a partial
trace of the degrees of freedom in the \emph{complement} of subsystem $A$:
\begin{equation}
    S = - \tr_A \left(\tr_B \ket{\psi}\bra{\psi}\right) \ln  \left(\tr_B \ket{\psi}\bra{\psi}\right).
    \label{eq:EE1def}
\end{equation}
As recently pointed out by Tarun Grover\cite{grover_certain_2014}, for MBL systems, SSA ensures that 
the disorder averaged von Neumann entanglement entropy at any energy density is concave, where
$\bar{S}(l,e)$ is averaged over a tiny energy density window and all disorder configurations.

Unfortunately, without disorder averaging, the curve $S(\ell)$ for individual eigenstates obeys no concavity
conditions showing essentially random behavior, which stems from the local entanglement structure
(\cf Fig. \ref{fig:averageEE} Top).

In this section, we show how to use SSA
to derive constraints on $S(\ell)$ for 
any \emph{individual state} in a periodic one-dimensional system 
where the von Neumann entanglement entropy is averaged over all cuts with subsystem size $\ell$.  This 
average over all subsystems of size $\ell$ is sufficient to restore concavity for individual states (see Eq. \eqref{eq:CAEE})
and puts constraints on the sign of the slope at different $\ell$ (see Eq. \eqref{eq:SCAEE}).
This can be used to identify whether individual states separately obey an `area law' or a `volume law'.
We note that the result in Ref.~\onlinecite{grover_certain_2014} for periodic systems is a direct corollary of this result as the sum of concave functions is concave. 

\begin{figure}[h]
	\centering
	\includegraphics[scale=0.5]{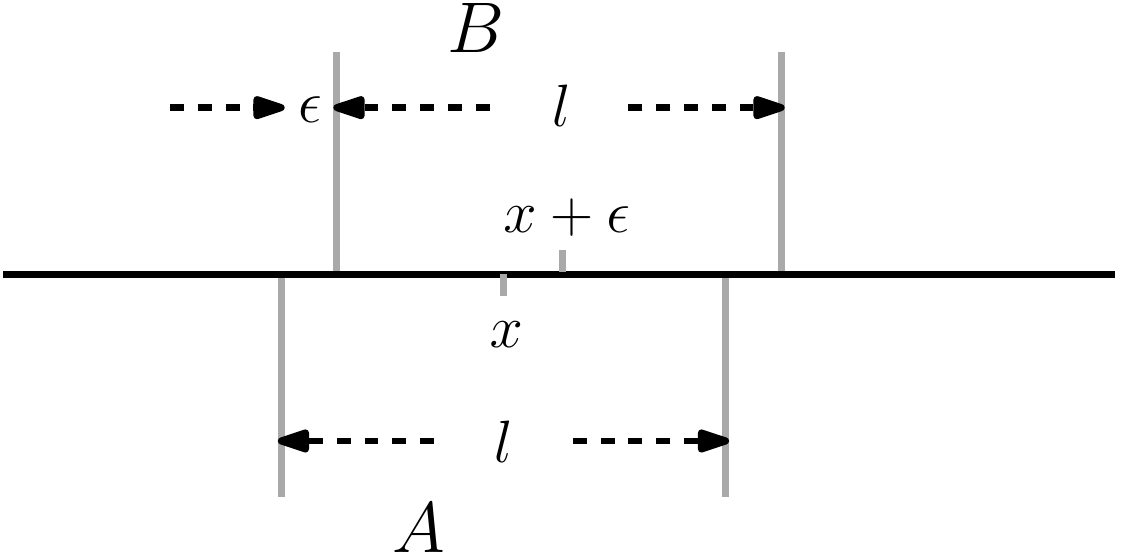}
	\caption{A typical arrangement of the subsystems to apply the strong subadditivity (SSA) inequality $S(A) + S(B) \ge S(A \cup B) + S(A \cap B)$. A continuous subsystem can be characterized sufficiently by its length $\ell$ and the position of its middle point $x$ in 1D.}
	\label{fig:SSA_region1}
\end{figure}

We focus on systems in 1D with periodic boundary conditions. $L$ denotes the length of the entire system. $\ell$ denotes the length of a subsystem, with $x$ as its center position. For the convenience of further discussion, we define the cut-averaged entanglement entropy (CAEE) as
\begin{equation}
	\bar{S}(\ell) \equiv \frac{1}{L} \int \limits_0^L \D x \ S(x,\ell), \tag{CAEE}
\label{eq:CAEE}
\end{equation}
and define the slope of the cut-averaged entanglement entropy (SCAEE) as
\begin{equation}
	\frac{\partial \bar{S}}{\partial \ell} \equiv \frac{1}{L} \frac{\partial }{\partial \ell} \int
    \limits_0^L \D x \ S(x,\ell). \tag{SCAEE}
\label{eq:SCAEE}
\end{equation}

For a density matrix of any quantum state, according to SSA, the von Neumann entanglement entropy obeys
\begin{equation}
S(A) + S(B) \ge S(A \cup B) + S(A \cap B).
\label{eq:SSAp}
\end{equation}
We apply this where $A$ and $B$ are subsystems of equal length, but slightly shifted apart as in Fig.
\ref{fig:SSA_region1}. Translational invariance is not assumed due to the presence of disorder and we
focus on single eigenstates of a particular disorder pattern from now on.

The inequality \eqref{eq:SSAp} is equivalent to
\begin{equation}\label{eq:SSA1_discrete}
S(x,\ell) + S(x+\epsilon,\ell) \ge S(x+\frac{\epsilon}{2},\ell+\epsilon) +
S(x+\frac{\epsilon}{2},\ell-\epsilon).
\end{equation}
When expanded to second order in $\epsilon$, it becomes
\begin{equation}\label{eq:SSA1}
\frac{\partial^2 S(x,\ell)}{\partial^2 \ell} \le \frac{1}{4} \frac{\partial^2 S(x,\ell)}{\partial^2 x}.
\end{equation}
For a system with periodic boundary conditions, by integrating the above equation over the entire system, it is easy to see that
\begin{equation}
    \label{eq:SSA_curvature}
	\frac{\partial^2 }{\partial^2 \ell} \bar{S}(\ell) \le 0,
\end{equation}
because the boundary terms at $x=0$ and $x=L$ cancel each other exactly. Eq. \eqref{eq:SSA_curvature} puts a constraint on the concavity of $\bar{S}(\ell)$ which holds for any given eigenstate of an arbitrary disorder configuration, because SSA is applicable to the density matrix of any quantum state.

\begin{figure}[h]
	\centering
	\includegraphics[scale=0.5]{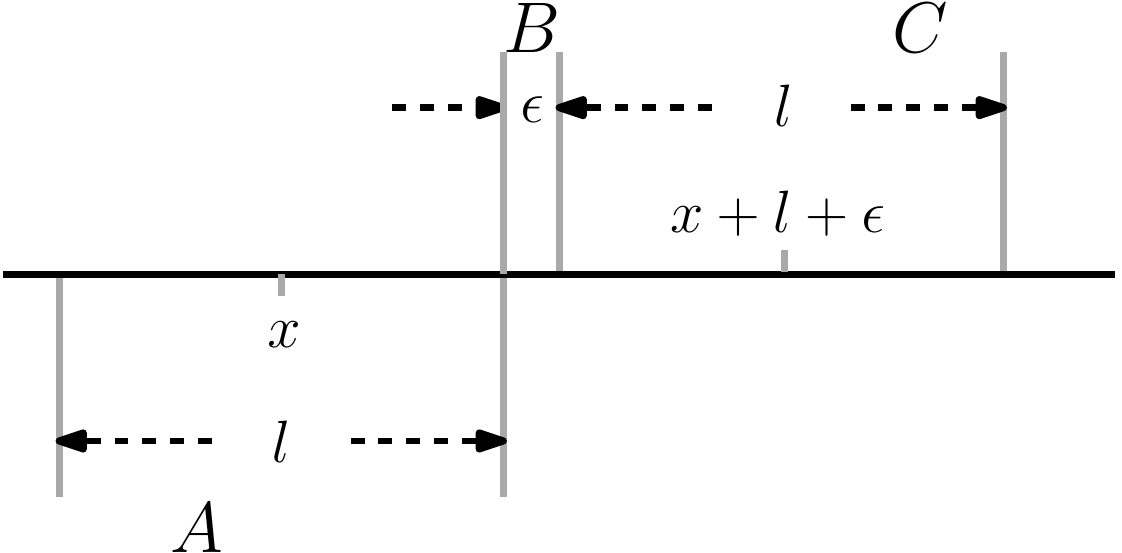}
	\caption{A typical arrangement of the subsystems to apply the strong subadditivity (SSA) inequality $S(A \cup B) + S(B \cup C) \ge S(A) + S(C)$. A and C are non-overlapping, which means $l<L/2$, where $L$ is the length of the entire periodic system.}
	\label{fig:SSA_region2}
\end{figure}

There exists an essentially equivalent constraint on the first derivative of $\bar{S}(\ell)$, based on an equivalent formulation of the SSA theorem as in Fig. \ref{fig:SSA_region2}.
\begin{equation}
S(A \cup B) + S(B \cup C) \ge S(A) + S(C),
\end{equation}
When written explicitly, it becomes
\begin{equation}\label{eq:SSA2_discrete}
S(x+\frac{\epsilon}{2},\ell+\epsilon) + S(x+\ell+\frac{\epsilon}{2},\ell+\epsilon) \ge S(x,\ell) +
S(x+\ell+\epsilon,\ell).
\end{equation}
One can expand it to the first order in $\epsilon$
\begin{equation}
\frac{\partial S(x,\ell)}{\partial \ell} + \frac{\partial S(x+\ell,\ell)}{\partial \ell} \ge
\frac{1}{2} \frac{\partial S(x+\ell,\ell)}{\partial x} - \frac{1}{2} \frac{\partial S(x,\ell)}{\partial x}.
\end{equation}
Again, with periodic boundary conditions, integrating the above equation over the entire system reduces to
\begin{equation}
	\label{eq:SSA2}
	\frac{\partial \bar{S}(\ell)}{\partial \ell} \ge 0.
\end{equation}
where $\ell< L/2$.

The two constraints \eqref{eq:SSA_curvature} and \eqref{eq:SSA2} imply that the cut
averaged entanglement entropy $\bar{S}(\ell)$ is a concave function of subsystem size with
positive slope for $0\leq \ell \leq L/2$ and negative slope for $L/2 \leq \ell \leq L$. Its
negative second derivative makes the slope of larger subsystems at most equal to or smaller than the
slope of smaller subsystems. It should be emphasized that these considerations are only valid for
the cut averaged entanglement entropy in periodic systems. Without this average, the `slope' of the
$S(\ell)$ curve can become negative even for $\ell<L/2$, which we argue to be a sign for localized
regions in the system (\cf Fig. \ref{fig:slopenomean}).

\section{Model and Method}\label{model_method}

We study the ``standard model'' of the MBL transition: The periodic random field Heisenberg
chain\cite{oganesyan_localization_2007,znidaric_many-body_2008,pal_many-body_2010,luca_ergodicity_2013,pekker_encoding_2014,luitz_many-body_2015,yu_finding_2015,serbyn_criterion_2015,bera_many-body_2015,luitz_extended_2016,singh_signatures_2016,pollmann_efficient_2015,khemani_obtaining_2015,lim_nature_2015,luitz_long_2016,bera_local_2016},
described by the Hamiltonian

\begin{equation}
    H= \sum_{i=1}^{L} \hat{\vec{S}}_i \cdot \hat{\vec{S}}_{i+1} + h_i \hat{S}_i^z, \,\,\,\,
    h_i\in[-h,h]; \,\, p(h_i)= \frac{1}{2h},   
    \label{eq:model}
\end{equation}
where the site dependent magnetic field $h_i$ is a uniform random number, coupling to $\hat{S}_i^z$.
In this system current evidence primarily suggests an  
MBL transition which occurs at a critical disorder strength $h_c\approx 3.7$ in the center of
the spectrum\cite{pal_many-body_2010,luitz_many-body_2015}, although the value of $h_c$ is not fully
settled\cite{devakul_early_2015}.

Interior eigenpairs are obtained using a
shift-invert technique, where the Hamiltonian is transformed to $\left(\mat{H}-\sigma\right)^{-1}$,
with a target energy $\sigma$ inside the spectrum. The transformed problem is then amenable to
standard Krylov space methods to obtain eigenpairs from the (upper and lower) edges of the transformed
spectrum, reducing the computational difficulty virtually to the problem of applying the inverse of
the shifted Hamiltonian to arbitrary vectors. This is a formidable task, due to the rapid growth of
the problem dimension and high density of states in the middle of the spectrum. 
Currently, no more than $L=22$ spins in
the $S_z=0$ sector of the random Heisenberg chain can be treated using the shift-invert methodology 
even when applied in a massively
parallel way\cite{luitz_many-body_2015}.
Throughout this work, we address eigenpairs in the center of the spectrum at fixed energy density
$\epsilon = (E-E_\text{min})/(E_\text{max} - E_\text{min})=0.5$, where the extensive target energy
$E$ in each disorder configuration is determined by the corresponding groundstate ($E_\text{min}$)
and antigroundstate ($E_\text{max}$) energies.

\section{Cut averaged entanglement entropy (CAEE)} \label{Cut_averaged_EE}

The considerations in Sec. \ref{sec:continuous} (and appendix \ref{sec:discrete}) represent a strong constraint on
the CAEE as a function of subsystem size. In particular, for $\ell \le L/2$, the
CAEE for larger subsystems always has to be larger than (or equal to) the ones for smaller
subsystems. Note that, while this property holds for individual disorder
realizations, the disorder average of the cut-averaged entanglement entropy is identical with the
disorder average of the standard (single cut) entanglement entropy.

The bottom panel of Figure \ref{fig:averageEE} shows typical CAEE curves as a function of
subsystem size $\ell$ obtained from single (mid-spectrum) eigenstates of the Hamiltonian at various
disorder strengths $h$. Due to the subadditivity constraints, the value of the CAEE at a given subsystem size has to be 
correlated with its values at other subsystem sizes.
The CAEE curve is mirror symmetric around $\ell=L/2$ and finite size effects seem to be strongest in the region of
the half cut, especially for the slope of the
curve. We therefore primarily focus on the quarter cut $\ell=L/4$, where these effects are much less
important. Notice that in the absence of cut averaging (\cf top panel of Figure
\ref{fig:averageEE}), the entanglement entropy as a function of subsystem size is
non-differentiable and therefore its slope is not well defined.

For each eigenstate, we calculate the reduced density matrix of all possible cuts of the system with
a fixed subsystem length $\ell$ and average the von Neumann entropy over them. This is repeated for all subsystem sizes
$\ell \in [1,L/2]$, yielding the CAEE $\bar{S}(\ell)$ as a function of the (integer)
subsystem size $\ell$. In the next step, we interpolate this set of points (extended down to
$\ell=0$ and up to $\ell=L$ by the symmetry $S_\ell=S_{L-\ell}$) by a cubic spline,
yielding a smooth and continuous function of $\bar{S}(\ell)$. The result for several typical states
is depicted by the lines connecting the points of Fig. \ref{fig:averageEE} for various values of the disorder strength. 
This allows us to
look at derivatives of the entanglement entropy curve at fixed ratio $\ell/L$,
which ensures the limit of an extensive subsystem size and can be estimated even for system sizes,
where $\ell/L$ is incommensurate with the lattice. We have checked that using discrete derivatives
(for the commensurate cases) yields virtually identical results and does not change the conclusions of
this work. The corresponding derivation of the bounds in the discrete case is provided in appendix
\ref{sec:discrete}.

\subsection{Cut averaged entanglement entropy slope (SCAEE)}

We calculate the CAEE $\bar{S}(\ell)$ as a function of subsystem size for all eigenstates at fixed energy density
$\epsilon=0.5$ and determine the SCAEE $\partial \bar{S}(\ell)/\partial \ell$. 
A fully ergodic eigenstate should have a SCAEE close to its maximal value: $\ln 2$.
On the other hand, if the state is fully localized, the slope should be close to zero.
In this section, we present detailed results on the SCAEE in different 
regimes; we consider its distribution over disorder realizations
considering the first and second moment of the distribution as well as the entire distribution itself.

\subsubsection{Mean slope}

\begin{figure}[h]
    \centering
    \includegraphics{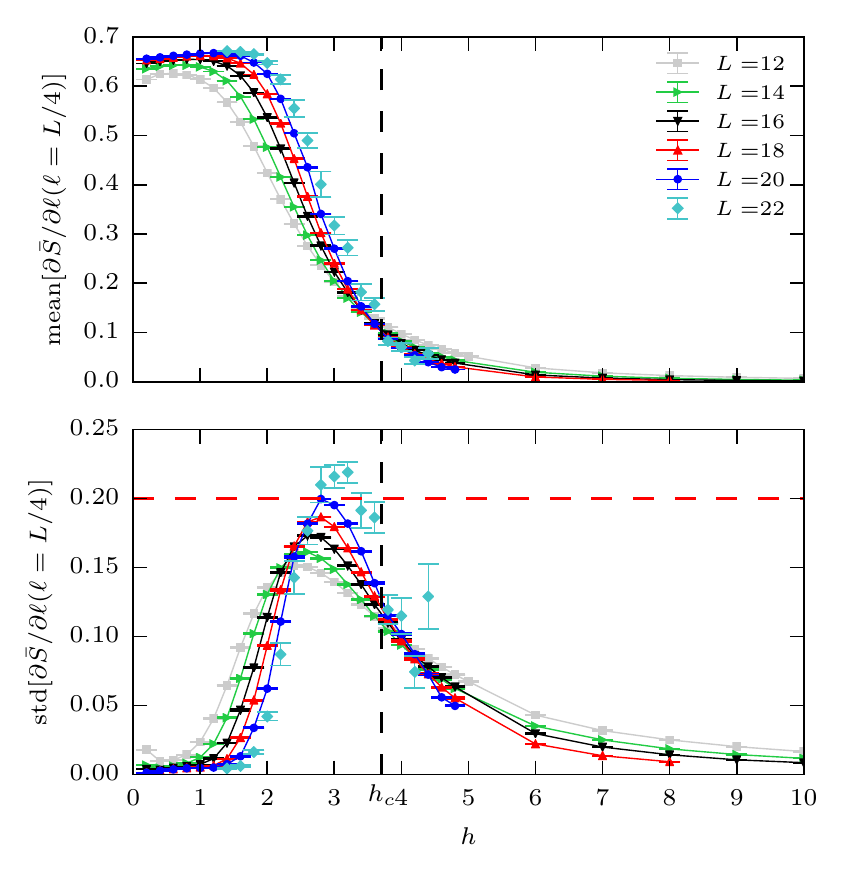}
    \caption{Disorder averaged mean entanglement entropy slope (top) and standard deviation of the
        entanglement entropy slope (bottom). The black dashed lines mark the rough intersecting positions
		of the curves of different system sizes. The red horizontal line for the bottom panel indicates the 
		standard deviation of a uniform (box) distribution on the interval $[0,\ln 2]$, given by $\sigma_\square 
		= \frac{\ln 2}{2 \sqrt{3}}$. 
        Note that the maximal variance of a bounded distribution on $[0,\ln 2]$ is given by a
        bimodal delta distribution, yielding $\sigma_= \frac{\ln 2}{2} \approx .34657$. The maximum
    can at most saturate at this value.}
    \label{fig:mean_slope}
\end{figure}

We begin with the mean of the SCAEE for subsystems of size
$\ell=L/4$ for different system sizes. The top panel of Fig. \ref{fig:mean_slope} shows the result
obtained from the disorder average over $\approx 10^3$ disorder configurations and $\approx 50$
eigenstates per realization as a function of disorder strength. The best estimate of the critical
point in the center of the spectrum from Ref. \onlinecite{luitz_many-body_2015} is indicated by the vertical dashed line.

\begin{figure}[h]
    \centering
    \includegraphics{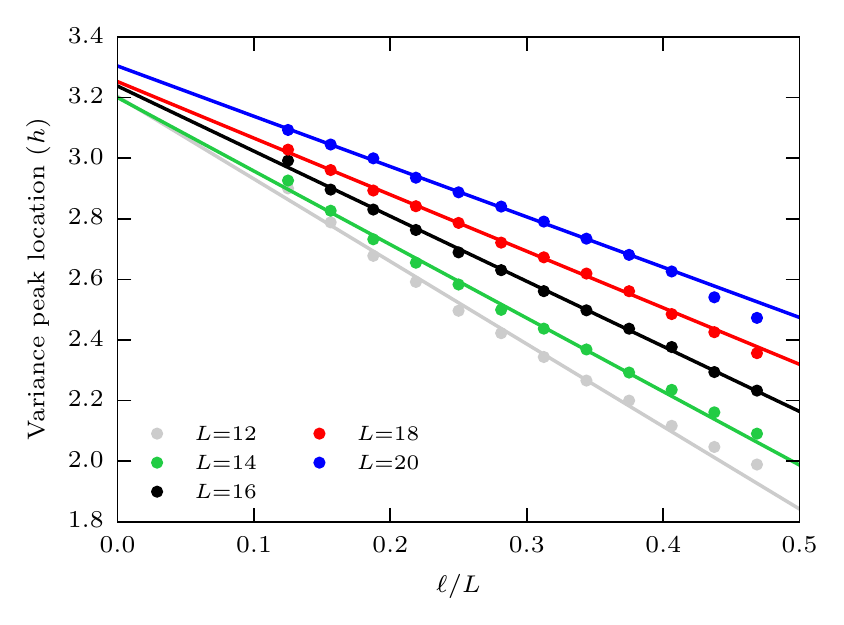}
    \caption{Variance peak locations at different subsystem ratio $\ell/L$ for various system sizes.
        The variance peak locations are extracted from standard deviation plots as in Fig.
        \ref{fig:mean_slope}, where the discrete $\bar{S}(\ell)$ are interpolated by cubic splines.
    The solid lines are linear fit curves to data points. The variance peak location (h) rises with
increasing system sizes. It is important to notice that for a particular $L$, the differences
between the peak locations at different ratios becomes smaller, as $L$ increases, which is reflected
by the decreasing slope of the lines connecting different subsystem sizes.}
    \label{fig:peak_location}
\end{figure}

At weak disorder, the slope gets closer to its maximal value of $\ln 2$ for increasing system size. In addition, 
at larger system sizes, the $\ln 2$ plateau extends to larger values of $h$ consistent with the idea of a sharp transition away from $\ln 2$ in the TDL. 
Near the critical point, the average slope decreases
rapidly to a smaller (near zero) value in the MBL phase.  As the system size grows, the slope 
outside the critical region decreases as system size increases. These
behaviors are fully consistent with the results obtained for other quantities and in particular the
entanglement entropy density $S/L$, studied in Ref. \onlinecite{luitz_many-body_2015}, a quantity
that is closely related to the SCAEE if averaged over disorder, as we shall see in Sec.
\ref{sec:corrslope}.   

In the critical region we see that the mean SCAEE for different system sizes generally intersect.
While finite size effects will cause this crossing to drift with system size, we may treat it as a 
rough estimate for the critical point and find it consistent with the critical $h_c$ estimated
by other approaches\cite{pal_many-body_2010,luitz_many-body_2015}.   

It is interesting to note that the mean SCAEE is 
approximately 0.1 at this crossing point.  Again, noting that there will be a drift of this value, 
a non-zero slope implies a non-area law value of the entanglement at the transition.

\subsubsection{Variance peak of the entanglement entropy slope}
\label{sec:variancePeak}
Recently, it has been discovered that the variance of the distribution of the entanglement entropy
in MBL systems exhibits a maximum close to the MBL
transition\cite{kjall_many-body_2014,luitz_many-body_2015,chen_many-body_2015}.
In Fig. \ref{fig:mean_slope}, we present results on the standard deviation of the SCAEE,
$\text{std}[ \partial \bar{S}/\partial \ell ]$, for subsystem sizes $\ell=L/4$.
This quantity also exhibits a maximum close to the transition and although the slope is an intensive
quantity, the variance peak seems to grow in amplitude for the accessible system sizes $L\leq22$.

Because the cut-averaged entanglement entropy slope is strictly bounded by $0\leq \partial
\bar{S}/\partial \ell \leq \ln 2$, the peak of the standard deviation can not grow infinitely and
can at most saturate\footnote{We have tried a tentative extrapolation of the peak height to the
thermodynamic limit using a polynomial ansatz as a function of $1/L$, pointing to a maximal
variance of $\ln 2/2$, although this extrapolation is not reliable, as the extrapolation typically
overshoots the theoretical maximum due to finite size effects.} 
at the maximal standard deviation given by a bimodal delta distribution $p(x) =
\frac{1}{2} \left(\delta(x) + \delta(x-\ln 2)\right)$.

We will discuss the origin of the variance peak of the entanglement entropy slope below, when we
present the full probability distributions of this quantity and show that it becomes bimodal close
to the transition. This is clear from the behavior of the standard deviation with system size in Fig.
\ref{fig:mean_slope}, which grows for the available size as a function of $L$ and exceeds the value
obtained for a box distribution for system sizes larger or equal than $L=20$. This bound is
important as it represents the \emph{largest} standard deviation of a unimodal distribution on
$[0,\ln2]$. Larger variances necessarily imply that the distribution has to be \emph{multimodal} and the
inspection of our histograms in Sec. \ref{sec:slopehisto} clearly points to a bimodal distribution
with maxima close to $0$ and $\ln 2$.

We attempted polynomial extrapolations of the maxima of the variance to the TDL. 
Given our system sizes and the lack of justification for the scaling ansatz, such an extrapolation
is obviously unreliable but the results are not inconsistent with the maxima reaching the maximal
possible variance ($\frac{\ln 2}{2}$) close to the critical point ($h_c=3.7(1)$).

In Fig. \ref{fig:mean_slope} we have chosen a subsystem of size $\ell=L/4$ to mitigate
strong finite size effects occurring at the half cut. The difference between the peak
locations of various subsystem ratios becomes smaller as system size grows (\cf Fig.
\ref{fig:peak_location}).
This suggests that the transition, as identified by the variance peak, occurs at the same value of $h$ for all subsystem sizes in the
TDL. This latter fact is a necessary condition if all states in the TDL have
linear (possibly zero) SCAEE for all subsystem sizes.

\subsubsection{Slope histogram}
\label{sec:slopehisto}

It is known from many studies
\cite{vosk_theory_2015,potter_universal_2015,bar_lev_absence_2015,agarwal_anomalous_2015,luitz_extended_2016,varma_energy_2015,gopalakrishnan_griffiths_2016,luitz_long_2016}
that the surrounding of the MBL transition is dominated by rare region effects, responsible for
subdiffusion\cite{bar_lev_absence_2015,agarwal_anomalous_2015,luitz_extended_2016,varma_energy_2015}
on the thermal side of the transition. This is reflected in pathological (non-Gaussian) features of
the probability distributions over disorder of various observables\cite{luitz_long_2016}. In
particular the entanglement entropy \cite{bauer_area_2013,lim_nature_2015,luitz_long_2016} develops
long tails down to zero entanglement, although the mean remains extensive and the weight of the
tails is exponentially suppressed. 

Here, we are interested in the SCAEE, which can be expected to
wash out some of the rare region effects, while capturing the dominant scaling behavior of the entanglement
entropy. As in the above discussion, we focus on extensive subsystems, where we expect
that the localization length $\xi$ will be overcome at subsystem system sizes where $\ell\gg \xi$,
thus providing a clear separation of dominant localized and delocalized behaviors for large system
sizes.

\begin{figure}[h]
    \centering
	\includegraphics{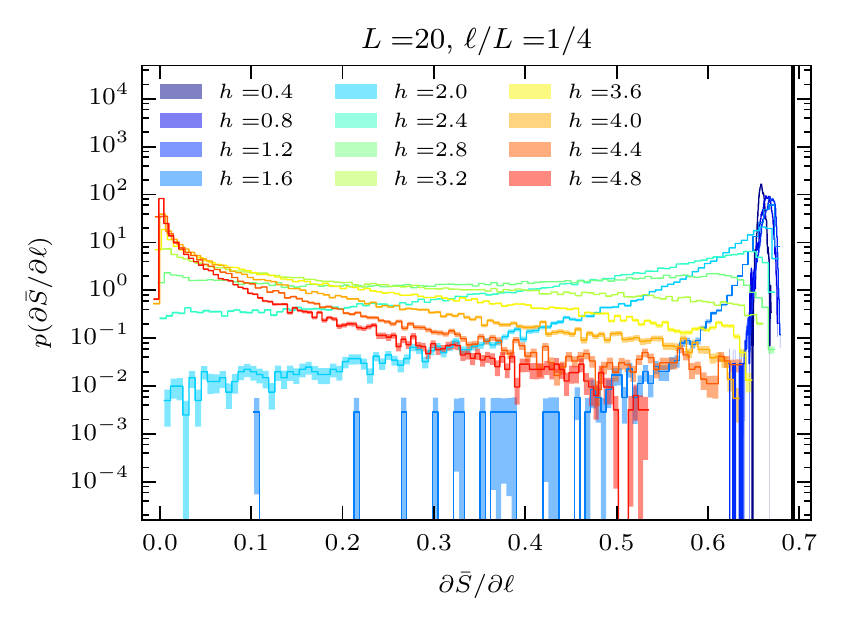}
    \caption{Probability distribution of SCAEE evaluated at $\ell = L/4$, for systems of size
    $L=20$ at various disorder strengths. The black vertical line indicates the maximal slope of $\ln{2}$.
	}
    \label{fig:eeoverview}
\end{figure}

\begin{figure}[h]
    \centering
	\includegraphics{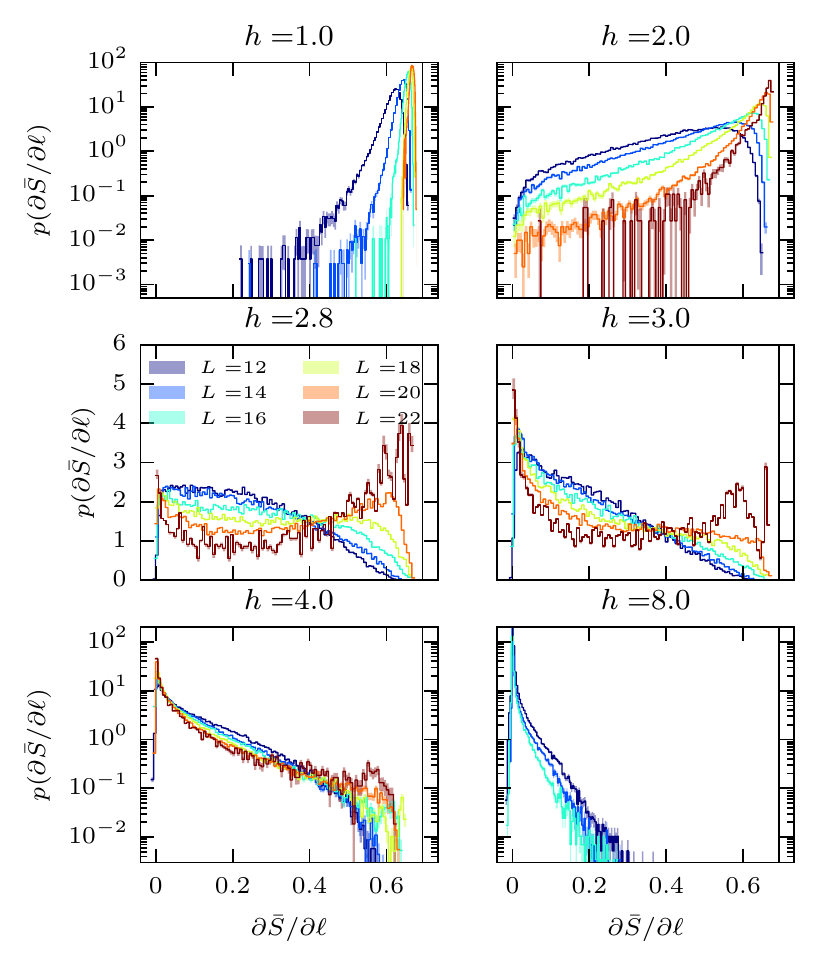}
    \caption{System size dependence of the probability distributions of SCAEE evaluated at $\ell/L=1/4$.
	At weak disorder ($h=1.$ and $h=2.$), the distribution approaches a Gaussian distribution for large system 
	sizes. For our finite systems, the zone showing critical behaviour is roughly at $h=3.0$ and drifts to larger 
	disorder strength for larger systems. We observe a clear signature of an emerging bimodal
    distribution. In 
	the MBL phase ($h=4$ and $h=8$), the distribution is again unimodal and sharply peaked at zero slope. 
	The black vertical line indicates the maximal slope of $\ln{2}$. Note that the panels in the
critical region are shown on a linear scale for clarity, while the other panels are on a logarithmic
scale to exhibit the tails.}
    \label{fig:eesize}
\end{figure}

    At weak disorder, the distribution of the SCAEE shown in Fig.
    \ref{fig:eeoverview} is close to a normal distribution with a mean close to the maximal slope at
    $\ln 2$, indicated by the vertical line. As also seen in the entanglement entropy presented in Ref.
    \onlinecite{luitz_long_2016}, there may be small deviations from the normal distribution, surviving in
    the thermodynamic limit but significant tails start to develop only at slightly larger disorder
    strengths around $h\gtrsim 0.8$. The weight in the tails reduces with system size as seen in
    Fig. \ref{fig:eesize} but the distributions remain generically non-Gaussian starting at
    relatively small disorder strengths of $h\approx 1$.

    Near
    the critical point (middle panels of Fig. \ref{fig:eesize}), the distribution becomes broad, giving rise to the maximal variance as
    discussed in the previous section. In fact, with growing system sizes, the distribution becomes
    increasingly bimodal, developing maxima close to zero (minimal) slope and $\ln 2$ (maximal) slope. 
    As the position
    of the variance peak (lower panel of Fig. \ref{fig:averageEE}) moves towards the critical point
    for large systems and becomes sharp, we expect a bimodal distribution of the entanglement slope
    to be characteristic for the MBL transition. We show in Fig. \ref{fig:averageEE} that the value
    of the variance at its maximum is already slightly larger than the value of a uniform
    distribution for systems of size $L=20$ and is definitely larger than the uniform variance for
    larger system sizes (with growing tendency of the peak height for larger system sizes, possibly
    up to saturation at the theoretical maximum of the standard deviation at $\sigma_\text{max} =
    \frac{\ln 2}{2\sqrt 3}$). This rules out the possibility of a flat or unimodal distribution, leaving as the only possibility consistent with our results for the shapes of
    the distribution a bimodal distribution with maxima close to the minimal and maximal slope.
    Whether the weight between these maxima vanishes completely in the thermodynamic limit can not be
    definitely answered from our finite size results but this scenario is consistent with our data.

    In the MBL phase, the maximum of the distribution of the SCAEE has
    clearly shifted towards very small slopes with an exponentially suppressed tail, extending up to
    the maximal slope. For the $h=8.0$ plot, one observes a small weight for negative slopes, which
    are an artifact of our spline interpolation: For very low entanglement entropies, the spline
    tends to become oscillatory and leads to slightly negative slopes. This is not a problem for
    larger entanglement entropies.

\begin{figure}[h]
    \centering
    \includegraphics{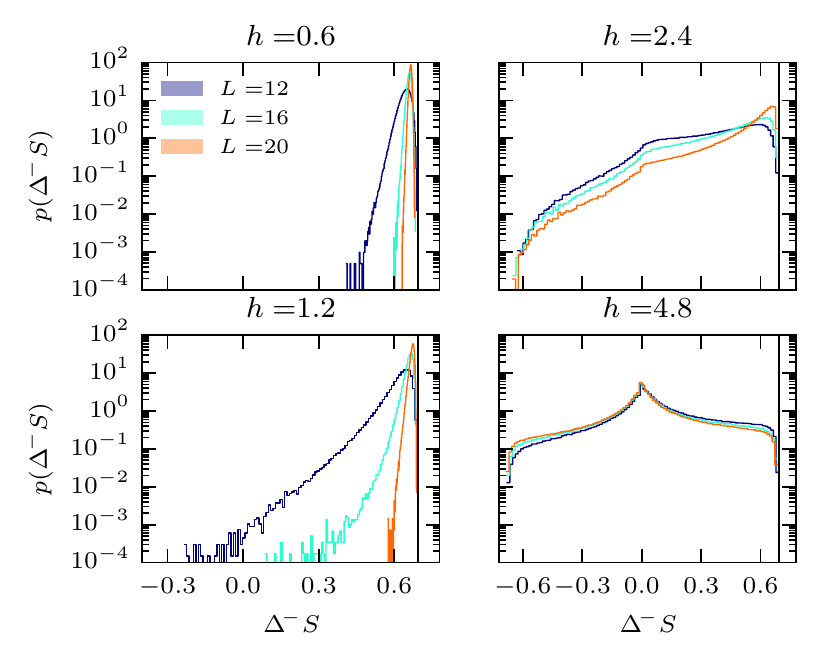}
    \caption{Discrete entanglement entropy ``slope" $\Delta^{\!-}S(L/4) = S(L/4)-S(L/4-1)$ of the raw
    entanglement entropy \emph{without averaging over all subsystem cuts}. The appearance of
    negative slopes is associated with localized regions\cite{luitz_long_2016}. 
	}
    \label{fig:slopenomean}
\end{figure}

    While the SCAEE captures the \emph{dominant} scaling behavior of the
    entanglement entropy, it has a tendency to hide the effect of rare (localized and ergodic)
    regions. Therefore, we also study the distribution of the discrete entanglement entropy slope
    $\Delta^{\!-}S(L/4)=S(L/4) - S(L/4-1)$ \emph{without averaging over cuts} in Fig.
    \ref{fig:slopenomean}. 
    The most striking difference to the cut averaged slope is the absence of the
    SSA constraints, allowing for a \emph{decrease} of the
    entanglement entropy with increasing system size, which can typically be expected if the
    boundary of the subsystem touches a localized part of the system as recently discussed in Ref.
    \onlinecite{luitz_long_2016}.

    At weak disorder strength ($h\lesssim 0.8$), the histogram of $\Delta^{\!-}S$ approaches 
    a Gaussian distribution for large system sizes, very similar to the cut averaged slope and no
    negative discrete slopes $\Delta^{\!-}S$ are observed. This changes significantly at
    intermediate disorder $h\gtrsim 2$, where more weight at negative discrete slopes is built up
    and a peak at 0 appears. We may speculate that this peak is caused by situations in
    which the changing subsystem boundary of the subsystems of length $L/4$ and $L/4-1$ both lie in
    a localized region, thus leading to a small change of the entanglement entropy and consequently
    $\Delta^{\!-}S=0$. This peak becomes dominant in the MBL phase at $h\gtrsim3.7$, where the
    distribution of $\Delta^{\!-}S$ becomes increasingly symmetric for larger system sizes with
    positive and negative discrete slopes being equally probable. 
    The behaviour of these distributions is in very good agreement with the picture that 
	rare localized regions of the system exist in the ergodic phase, while in the MBL phase the
    role of localized and delocalized regions switches and the latter become rare.

\subsubsection{Variance peak of the entanglement entropy slope of individual disorder realizations}

In Sec. \ref{sec:slopehisto} we find a bimodal distribution of the SCAEE when 
sampled over eigenstates and disorder realizations. Here, we consider this distribution
over single disorder realizations. In particular, we show that (i) eigenstates within a 
single sample can (but do not always) show bimodal behavior of the SCAEE (ii)
partially contribute to the variance of the SCAEE seen in Sec. \ref{sec:variancePeak} and (iii)
have values of the SCAEE which are not independent of each other.  
This means that within \emph{one disorder realization} some eigenstates show area
law scaling of the entanglement entropy, while others scale by a volume law, at fixed \emph{energy
density}. 

There are two extreme limits that can be considered in understanding the variance of the SCAEE.  
In one case, each disorder realization could individually contain eigenstates that all have  (near) maximal or (near) zero SCAEE and
hence individually have (near) zero variance.  Then the entire variance results from a distribution over different 
disorder realizations.
Alternatively, every eigenstate in a given disorder realization could be uncorrelated with each other.  We can emulate this case
by independently sampling states from the middle panels of fig.~\ref{fig:eesize}.  Sampling groups of 50 eigenstates
at $L=20$ and $h=2.8$ we see that the standard deviation of each group of 50 samples is a tight Gaussian centered around
$0.2$.  We find our data is consistent with neither extreme. Instead, both effects seem to be relevant
for the overall variance in Fig. \ref{fig:mean_slope}.

\begin{figure}[h]
    \centering
    \includegraphics{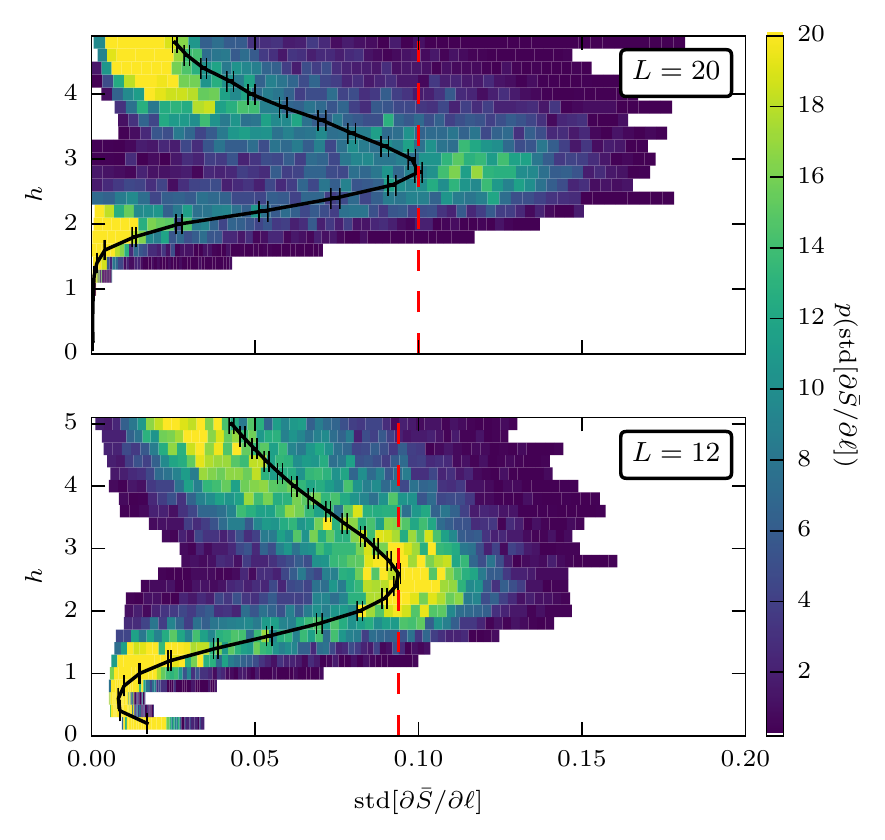}
    \caption{Probability density of the standard deviation of the SCAEE 
        $\partial \bar{S}/\partial \ell$ of eigenstates from the middle of the spectrum
        ($\epsilon=0.5$) of \emph{one disorder realization}
    with $\ell=L/4$. For each value of the disorder strength $h$, the distribution is sampled from
$\approx1000$ disorder realizations. Black curves indicate the mean of the distribution. Red dashed lines 
mark the largest standard deviations of the mean curves. A comparison of the results for system sizes $L=20$ and 
$L=12$ shows that for larger systems the states close to the critical point of \emph{one disorder realization} 
become more diverse.
}
    \label{fig:stdcolor}
\end{figure}

\begin{figure}[h]
    \centering
    \includegraphics{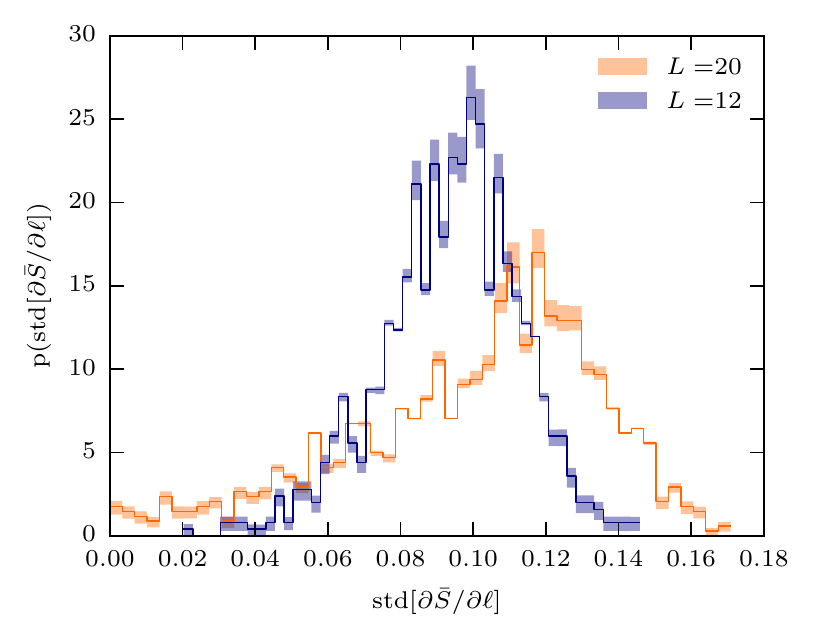}
    \caption{Histograms of horizontal slices ($h=2.6$ for $L=12$ and $h=2.8$ for $L=20$) through the
        mean curves' largest standard deviation points of Fig. \ref{fig:stdcolor}. In comparison,
    the one disorder realization distribution for a larger system size has more weight shifted
towards higher standard deviation, which indicates a higher likelihood of finding bimodality in a single sample.}
    \label{fig:std_hist}
\end{figure}

In order to understand if the large variance of the entanglement entropy and its slope can also be
(partly) created by single disorder realizations, we calculate the standard deviation of the SCAEE
\emph{for single disorder realizations} at fixed energy density
$\epsilon=0.5$ from approximately $50$ eigenstates per
disorder configuration.
Fig. \ref{fig:stdcolor} shows a two dimensional histogram of the per-sample standard deviation of
the SCAEE $\text{std}[\partial \bar{S}(L/4)/\partial \ell]$ for different disorder
strengths and system sizes $L=12$ and $L=20$. Most disorder realizations show a
larger variance of the SCAEE close to the critical point at $L=12$ and the disorder
averaged standard deviation (mean of the histogram) shown by the black line looks very similar to
the overall variance peak shown in Fig. \ref{fig:mean_slope}. For larger systems, we see the
following two trends in the critical regime (\cf Fig. \ref{fig:std_hist} for a cut
through Fig. \ref{fig:stdcolor} in the critical regime).  
First, the distributions over a single sample span a wider range of standard deviations as the  
system size grows.  There are therefore disorder realizations for which there is essentially no variance
and disorder realizations for which there is large variance among the eigenstates.  This is the opposite of
what we would expect if these eigenstates were independent, implying significant correlation between them.
Secondly, both the average per-sample standard deviation is larger
and the plurality of samples lie at higher standard deviations. This can be seen in Fig.~\ref{fig:stdcolor} where
the red dashed lines correspond to the means of the distributions at their
maxima. Note that the maximum of the distribution is actually larger than the mean close to the MBL 
transition, showing that typical realizations exhibit a mix of volume and area law states
(increasingly diverse for larger systems).
 
\begin{figure}[h]
    \centering
    \includegraphics{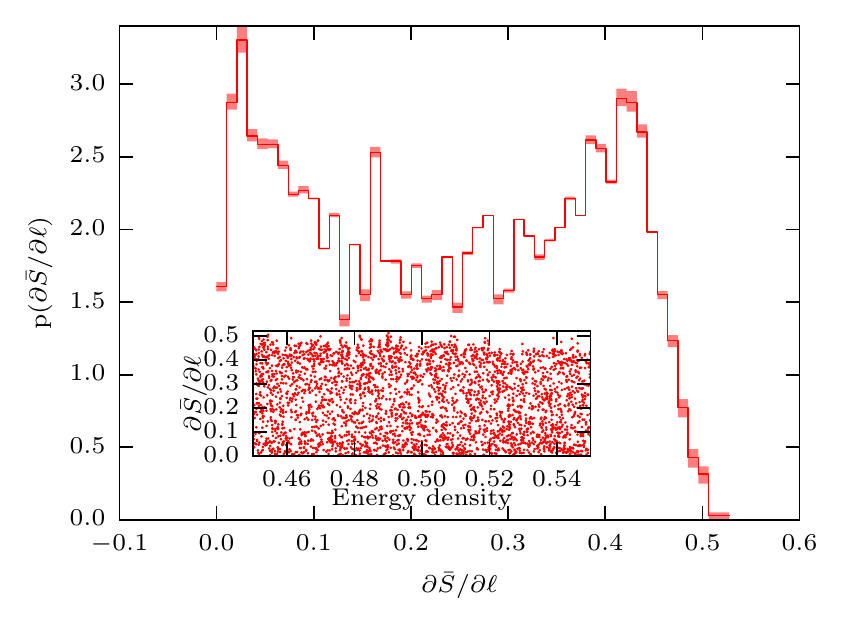}
    \caption{SCAEE distribution of a single disorder realization that has the largest standard
        deviation among the 500 samples at $L=16$. Each sample contains 6000 eigenstates, with
        disorder strength $h$ corresponding to the variance peak location of SCAEE evaluated at
        $\ell/L=1/4$ as in Fig. \ref{fig:mean_slope}. The inset shows SCAEE vs. energy density of
    each eigenstate of this disorder sample. Bimodality is clearly visible, and not likely due to the curvature of mobility edge.}
    \label{fig:single_sample_std_hist}
\end{figure}

 If these trends continue this means that the bimodality of the SCAEE distribution would arise at a single
disorder realization level, and become more and more common for increasing system
sizes.
To verify that bimodality is present for single disorder realizations, we numerically analyzed
500 disorder realizations of system size $L=16$ and a disorder strength $h=2.69$ that corresponds to the
variance peak location of the SCAEE at $\ell/L=1/4$. Each sample has about 6000 eigenstates within
the energy density window of $[0.45,0.55]$. Among the 500 samples, the one with the largest standard
deviation is shown in Fig. \ref{fig:single_sample_std_hist}. The distribution of the SCAEE for this
sample is strikingly bimodal. Note that this is not a result of the curvature of the mobility edge,
because the distribution of the SCAEE has no visible dependence on energy density as seen in the
inset of Fig. \ref{fig:single_sample_std_hist}. Especially, for this sample, the distributions of
the SCAEE of the eigenstates within energy density windows of $[0.45,0.50]$, $[0.50,0.55]$, and
$[0.45,0.475]\cup[0.525,0.55]$ are all bimodal and very similar to each other. Therefore, bimodality
does indeed seem to be a generic feature of the SCAEE of individual disorder realizations.

We speculate that the mechanism for this bimodality may be caused by a mix of quasi-local and extended $\tau$
operators\cite{serbyn_local_2013,imbrie_many-body_2014} at the transition, where volume law states correspond to corresponding occupied extended orbitals and area law states are given by an occupation of only localized orbitals.

\subsection{Correlations between the entanglement entropy and its derivatives}
\label{sec:corrslope}

In this section we show the nearly complete correlation between the CAEE density $\bar{S}/L$ and
the SCAEE $\partial \bar{S}/\partial \ell$ and between $\partial \bar{S}/\partial \ell$ and $\partial^2 \bar{S}/\partial \ell^2$ 
at a fixed subsystem ratio $\ell/L$ (which we choose as $1/4$), for various system sizes.  
These correlations are particularly compelling in the transition region, where, for finite system sizes, all 
accessible entropy densities are present (due to the wide and even bimodal distributions, depending
on system size). This suggests that in the approach to the TDL there is a universal one-parameter family of curves
$\bar{S}/L$ in the transition region parameterized by (for example) the value of the $\bar{S}/L$ at any fixed $\ell/L$.

\subsubsection{Correlation between $\bar{S}/L$ and $\partial \bar{S}/\partial \ell$}

\begin{figure}[h]
    \centering
	\includegraphics{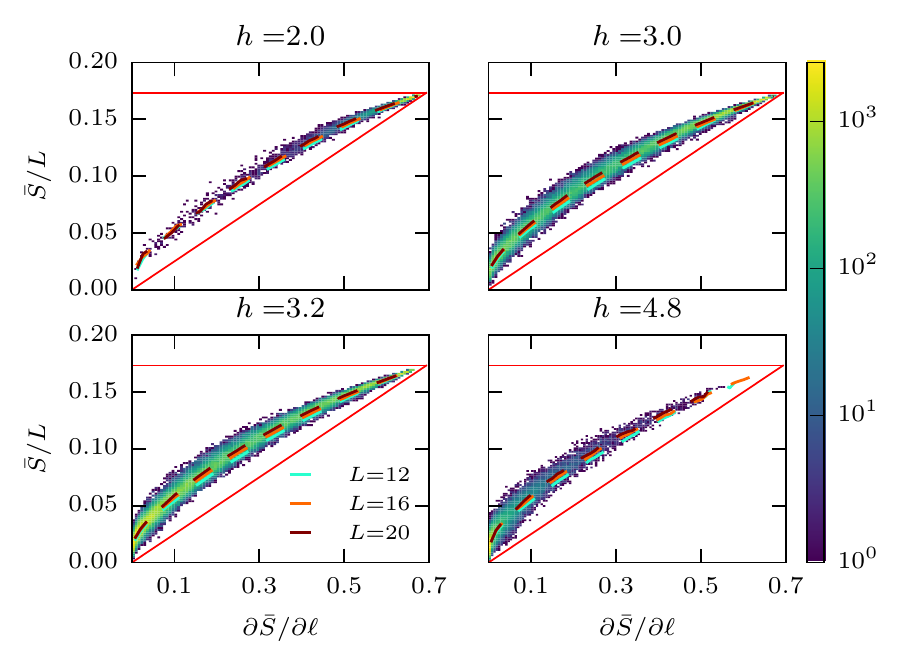}
    \caption{Two dimensional histogram of CAEE vs. SCAEE for systems of size
    $L=20$ and $\ell=L/4$, at disorder strength $h=2.0,3.0,3.2,4.0$. The mean curves for $L=12,16$ are also shown
    in the figure. The color bar indicates the bin counts on a logarithmic scale. The red straight lines indicate the upper ($\bar{S}/L \le \ln{2}/4$) and lower ($\bar{S}/L \ge \frac{1}{4} \partial \bar{S}/\partial \ell$) bounds of
    the entanglement entropy. The mean curves are nearly converged, indicating possible universal behavior.
	}
    \label{fig:2dhisto_deriv1}
\end{figure}

We show two dimensional histograms of $\bar{S}/L$ vs. $\partial \bar{S}/\partial \ell$ at various disorder strengths in Fig. \ref{fig:2dhisto_deriv1} for systems of size $L=20$, together with the mean curves for $L=12,16,20$ on the same plot. The color scale is logarithmic in the probability density.

The red lines are upper and lower bounds of $\bar{S}/L$ as a function of $\partial \bar{S}/\partial
\ell$, that can be derived from the SSA constraints: Because of the SSA constraint and the fact that
the slope is bounded from above by $\ln{2}$, we have
\begin{equation}
	\int_0^{L/4} \frac{\partial \bar{S}(L/4)}{\partial \ell} \D\ell \le \int_0^{L/4} \frac{\partial
        \bar{S}(\ell)}{\partial \ell} \D \ell \le \int_0^{L/4} \ln{2} \ \D \ell
\end{equation}
which reduces to
\begin{equation}
	\frac{1}{4} \frac{\partial \bar{S}(L/4)}{\partial \ell} \le \frac{\bar{S}(L/4)}{L} \le \frac{\ln{2}}{4}
\end{equation}

At small disorder strength ($h=2.0$), substantial weight is centered around $\partial \bar{S}/\partial \ell = \ln{2}$ and $\bar{S}/L=\ln(2)/4$, indicating volume law entanglement. A light tail extending all the way to the MBL (low entanglement and low slope) region is visible.

At large disorder strength ($h=4.8$), the weight is primarily around $\partial \bar{S}/\partial \ell = 0$ and $\bar{S}/L=0$, indicating area law entanglement. Again, a light tail extends to the ergodic region. At intermediate disorder strengths, for $L=20$, the weight spans from the ergodic to the MBL side. 

Notice, that there appears to be significant correlation between the CAEE and SCAEE and that the mean of these distributions 
seems to be largely independent of system size (at least from $L=12$ to $L=20$).  In fact, even for different disorder distributions
the banana-shaped histograms are located in the same region. 

Besides, as dicussed in Sec. \ref{sec:std_correlations}, we show that the width of the `banana' becomes narrower and more centered
around the mean values for larger system sizes.

All these points above are strongly suggesting the existence a one-parameter
family of curves of $\bar{S}/L$ in the transition region as a function of $\partial \bar{S}/\partial
\ell$ in the approach to the TDL. We also observed that the weight in the middle of the `banana'
becomes diminished with increasing system sizes, consistent with the previously discussed bimodality.

\subsubsection{Correlation between $\partial \bar{S}/\partial \ell$ and $L(\partial^2 \bar{S}/\partial \ell^2)$}

\begin{figure}[h]
    \centering
	\includegraphics{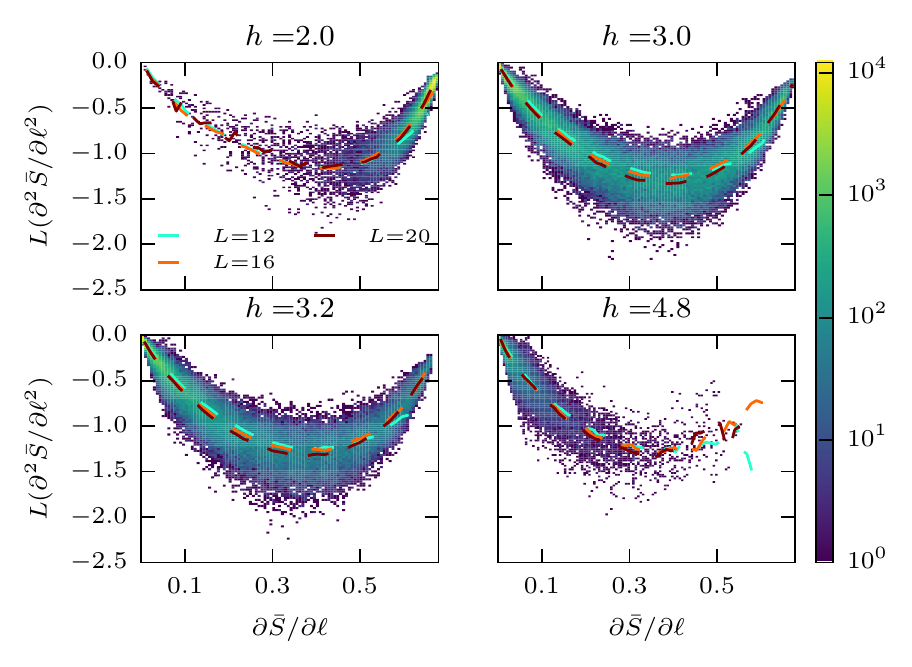}
    \caption{Two dimensional histogram of 2nd order derivative of CAEE vs. SCAEE for systems of size
    $L=20$ and $\ell=L/4$, at $h=2.0,3.0,3.2,4.0$. The mean curves for $L=12,16$ are also shown
	on the graphs. The color bar indicates the bin counts at a log scale. The mean curves are nearly converged, indicating possible universal behavior.}
    \label{fig:2dhisto_deriv2}
\end{figure}

We also consider the correlations between $\partial \bar{S}/\partial \ell$ and $L(\partial^2 \bar{S}/\partial \ell^2)$ again finding significant  correlation between them. 
The two dimensional histograms of $\partial \bar{S}/\partial \ell$ vs. $L(\partial^2 \bar{S}/\partial \ell^2)$ at various disorder strengths are shown in Fig. \ref{fig:2dhisto_deriv2} for systems of size $L=20$, together with the mean curves for $L=12,16,20$ on the same plot.

Similarly, one can make the following observations. At small disorder strength, the distribution is centered around $\partial \bar{S}/\partial \ell = \ln{2}$ and $L(\partial^2 \bar{S}/\partial \ell^2)=0$, indicating volume law entanglement entropy. At large disorder strength, the distribution is centered around $\partial \bar{S}/\partial \ell = 0$ and $L(\partial^2 \bar{S}/\partial \ell^2)=0$, indicating area law entanglement. At intermediate disorder strengths, the weights of the histograms span the high entanglement and low entanglement regions.  In addition, the lack of weight with 
concavity zero at values distant from zero or $\ln(2)$ slope indicates that we never see volume law (strictly linear)
curves with non-thermal (non-$\ln(2)$) or non-zero slope.
Again, we find that curves of system sizes between $L=12$ and $L=20$ have nearly identical means, and the width of the `U' shaped distribution becomes narrower and more centered around the mean values for larger system sizes, as dicussed in Sec. \ref{sec:std_correlations}.

\section{Conclusion}

We have presented an analysis of the scaling of the entanglement entropy up to the largest
accessible system sizes in state of the art exact diagonalization and introduced the cut averaged
entanglement entropy (CAEE) as well as its derivatives (SCAEE, \etc) with respect to subsystem size $\ell$ in periodic
chains. We anticipate that the CAEE and SCAEE will be useful quantities to study
the scaling behavior of the entanglement entropy 
of \emph{single eigenstates} in other inhomogeneous systems.

We find that the slope of the cut averaged entanglement entropy (SCAEE)
reproduces perfectly the volume law to area law transition associated with the MBL transition. 
We have also studied the slope
without performing the cut average and observe increasing weight at negative slopes, which we argue
to be associated with rare localized regions, which appear already in the ergodic phase and are
believed to be responsible for subdiffusion.

More interestingly, the probability distribution of the cut averaged entanglement entropy slope
becomes sharply bimodal for large system sizes in the transition region, leading to eigenstates
which have near zero and near maximal slope. The variance of this distribution in the critical
regime grows with system size, already exceeding the largest variance possible for a unimodal distribution
at $L=20$. In addition, we find that the scaling of
the CAEE is mostly linear in subsystem size at the variance peak with either maximal
($\ln 2$) or minimal ($0$) slope, but no state displays an intermediate value of the SCAEE and zero
curvature.
At the variance peak this implies that the scaling of the disorder averaged entanglement entropy is
a volume law with a coefficient below its thermal value.
We want to emphasize that due to the bimodal distribution, almost no state will show this behavior.

An extrapolation of the current trend is not inconsistent with the scenario that the
variance peak would reach the largest possible variance in the TDL, which would lead to a coefficient of the volume
law  at half the thermal value of $\ln(2)$ (in the disorder average).  

An important finding of our work is that this mixture of states is intrinsic and
occurs in single disorder realizations. Interestingly enough, it seems that there is significant 
correlation between eigenstates in a disorder realization and some such realizations actually have 
essentially no variance in their SCAEE of states within a tiny energy window. This mixture of states is potentially connected to the important open question on the
nature of the local integrals of motion at the transition. 

In the final part of this paper, we discussed the correlation between the cut averaged entanglement
entropy slope and the entropy itself and find that for large systems the value of the slope is a
strong predictor of the value of the entropy suggesting a one-parameter family of curves for the CAEE
in the critical region in the approach to TDL. In particular, this
observation allows us to draw conclusions on the behavior of the cut averaged entanglement entropy itself, which has to become bimodal, just as its slope. 

On a more speculative note, the observed bimodal features of the
distributions could be connected to the strong fluctuations of the entanglement entropy at the
critical point in RG calculations \cite{vosk_theory_2015,pekker_hilbert-glass_2014} and possibly to multifractal
features\cite{monthus_many-body-localization_2016}.

\begin{acknowledgments}
    DJL is very grateful to F. Alet, N. Laflorencie and Y. Bar Lev for many inspiring discussions and fruitful
    collaborations.  BKC acknowledges helful discussions with Anushya Chandran, Shivaji Sondhi, David Huse, 
    Vedika Kehmani, and David Pekker. 
    We thank F. Alet, Y. Bar Lev, D. Huse, N. Laflorencie, V. Khemani and C. Monthus for their
    helpful comments on the manuscript.
    A significant part of the results presented here has been obtained using computational
    resources of CALMIP (grant 2015-P0677) and the LPT at Universit\'e Paul Sabatier at Toulouse, France.

    This work was supported in part by the Gordon and Betty Moore Foundation's EPiQS Initiative
    through Grant No. GBMF4305 at the University of Illinois and the French ANR program
    ANR-11-IS04-005-01.
    Our code is partly based on the PETSc~\cite{petsc-web-page,petsc-user-ref,petsc-efficient}, 
    SLEPc~\cite{hernandez_slepc:_2005} and MUMPS\cite{MUMPS1,MUMPS2} libraries.
	
    This research is part of the Blue Waters sustained-petascale
    computing project, which is
    supported by the National Science Foundation (awards OCI-0725070
    and ACI-1238993) and the State
    of Illinois. Blue Waters is a joint effort of the University
    of Illinois at Urbana-Champaign and
    its National Center for Supercomputing Applications.
	
	BKC and XY acknowledge support from DOE, grant SciDAC FG02-12ER46875.
\end{acknowledgments}

\appendix

\section{Lattice version of SSA constraint}
\label{sec:discrete}
On discrete lattices, the continuum formulation of Sec. \ref{sec:continuous} may seem artificial,
however a completely analogous derivation can be obtained.
In this scenario, the $x$ and $\epsilon$ in Eq. \eqref{eq:SSA1_discrete} and Eq. \eqref{eq:SSA2_discrete} are
discrete numbers and in particular integers in units of the lattice constant.

Summing over all $x$ on a lattice for Eq. \eqref{eq:SSA1_discrete} and Eq. \eqref{eq:SSA2_discrete}, one easily obtains
\begin{equation}
2\bar{S}(\ell) \ge \bar{S}(\ell+\epsilon) + \bar{S}(\ell-\epsilon),
\end{equation}
and
\begin{equation}
2\bar{S}(\ell+\epsilon) \ge 2\bar{S}(\ell).
\end{equation}

Usually, one defines first and second order discrete derivatives with respect to $\ell$ as follows.
\begin{equation}
\Delta^{+} \bar{S}(\ell) \equiv \frac{\bar{S}(\ell+\epsilon) - \bar{S}(\ell)}{\epsilon}.
\end{equation}

\begin{equation}
\Delta \bar{S}(\ell) \equiv \frac{\bar{S}(\ell+\epsilon) - \bar{S}(\ell-\epsilon)}{2\epsilon}.
\end{equation}

\begin{equation}
\Delta^{-} \bar{S}(\ell) \equiv \frac{\bar{S}(\ell) - \bar{S}(\ell-\epsilon)}{\epsilon}.
\end{equation}

\begin{equation}
\Delta^2 \bar{S}(\ell) \equiv \frac{\bar{S}(\ell+\epsilon) + \bar{S}(\ell-\epsilon) -
2\bar{S}(\ell)}{\epsilon^2}.
\end{equation}

Then we have the following inequalities.
\begin{equation}
\Delta^2 \bar{S}(\ell) \le 0.
\end{equation}

\begin{equation}
\Delta^{+} \bar{S}(\ell) \ge 0, \ \Delta \bar{S}(\ell) \ge 0,\ \Delta^{-} \bar{S}(\ell) \ge 0.
\end{equation}

Also, it is straightforward that
\begin{equation}
\Delta \bar{S}(\ell) + \frac{\epsilon}{2} \Delta^2 \bar{S}(\ell) = \Delta^{+} \bar{S}(\ell).
\end{equation}

\begin{equation}
\Delta \bar{S}(\ell) - \frac{\epsilon}{2} \Delta^2 \bar{S}(\ell) = \Delta^{-} \bar{S}(\ell).
\end{equation}

And we can get a bound
\begin{equation}
\Delta \bar{S}(\ell) + \frac{\epsilon}{2} \Delta^2 \bar{S}(\ell) \ge 0.
\end{equation}

The inequalities Eq. (\ref{eq:SSA1}) and Eq. (\ref{eq:SSA2}) can be easily extended to higher
dimensions, and are fulfilled even for a single eigenstate. Importantly, these two equations imply
that $\bar{S}(\ell)$ is a smooth and monotonic function of subsystem size,
which is advantageous to study the volume law to area law transition of MBL
systems numerically.

In the present work, however, we use the continuum version of the SSA constraints by an analytic
continuation of the $\bar{S}(\ell)$ curve, because it allows us to study incommensurate subsystem
sizes. This is important due to the serious constraints in available system sizes from exact
diagonalization.

\section{Standard deviations of the correlations distributions}
\label{sec:std_correlations}
In Section \ref{sec:corrslope}, we dicussed the correlations between the CAEE density $\bar{S}/L$ and
the SCAEE $\partial \bar{S}/\partial \ell$ and between $\partial \bar{S}/\partial \ell$ and $\partial^2 \bar{S}/\partial \ell^2$ 
at a fixed subsystem ratio $\ell/L$ (which we choose as $1/4$), as shown in Fig \ref{fig:2dhisto_deriv1} and \ref{fig:2dhisto_deriv2}.
Here we further examine these distributions to establish the standard deviation, or spread, of each distribution around its mean, 
finding it narrower as the system size grows. 

\begin{figure}[h]
    \centering
	\includegraphics{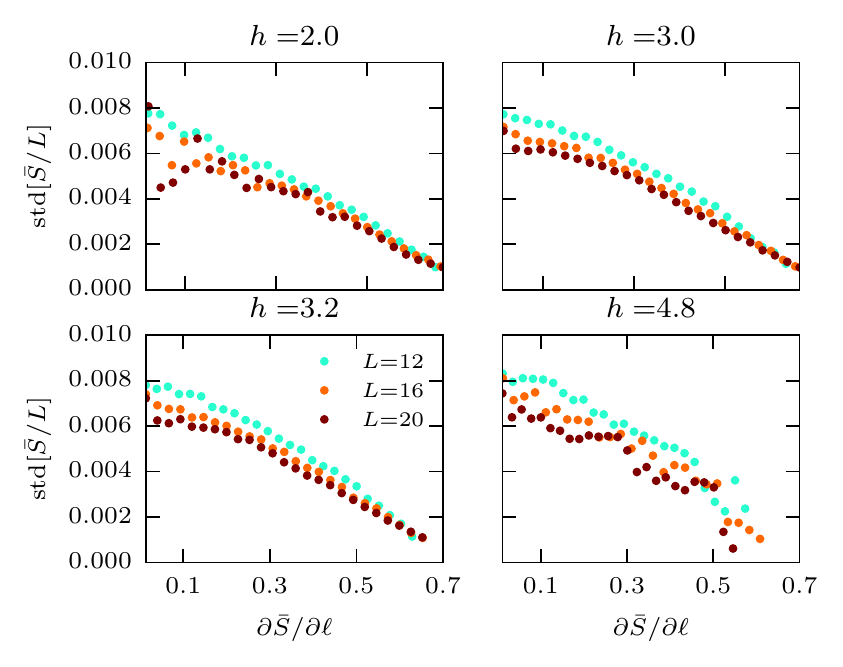}
    \caption{Standard deviation of vertical slices of CAEE vs. SCAEE histograms as in Fig. 
	\ref{fig:2dhisto_deriv1}, for systems of size
    $L=20$ and $\ell=L/4$, at $h=2.0,3.0,3.2,4.0$. The standard deviations decreases with system size.
	The 3rd and the 4th cumulants (not included) of the vertical slices of the histograms
	shows that the vertical distributions become more Gaussian with increasing system size.}
    \label{fig:2dhisto_deriv1_var}
\end{figure}

\begin{figure}[h]
    \centering
	\includegraphics{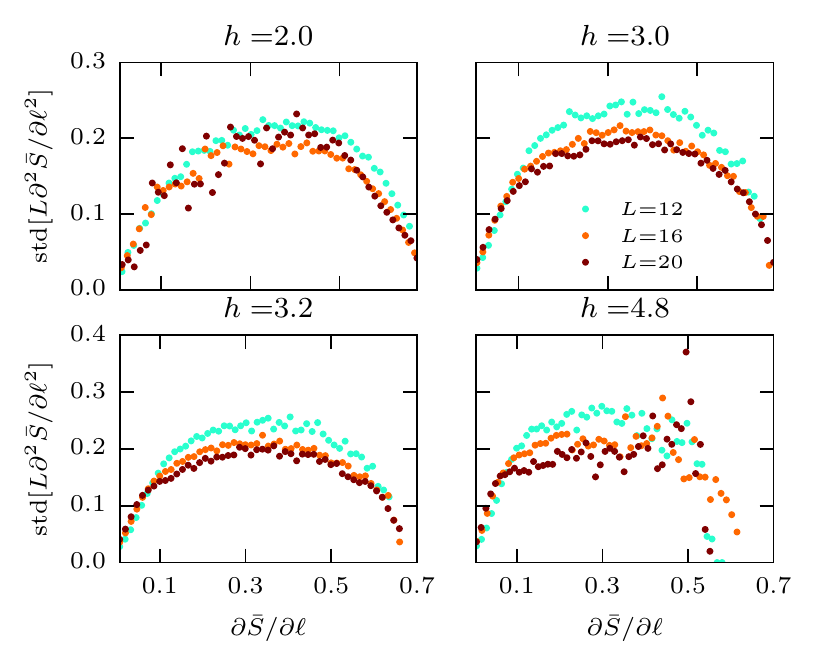}
    \caption{Standard deviation of vertical slices of 2nd order derivative of CAEE vs. SCAEE histograms as in 
	Fig. \ref{fig:2dhisto_deriv2}, for systems of size
    $L=20$ and $\ell=L/4$, at $h=2.0,3.0,3.2,4.0$. The standard deviations decreases with system size.
	The 3rd and the 4th cumulants (not included) of the vertical slices of the histograms
	shows that the vertical distributions become more Gaussian with increasing system size.}
    \label{fig:2dhisto_deriv2_var}
\end{figure}

For the 2D histograms of the CAEE density $\bar{S}/L$ and the SCAEE $\partial \bar{S}/\partial \ell$, we consider a slice of the two dimensional histograms vertically around a fixed value of $\partial
\bar{S}/\partial \ell$. This vertical slice gives a distribution of different $\bar{S}/L$ values at
some fixed value of $\partial \bar{S}/\partial \ell$. We calculated the third and fourth cumulants
of these distributions, and find their absolute values to be less then  $10^{-6}$  for $L=12$ and
even smaller for larger $L$; therefore these vertical slices are very likely Gaussian with well behaved mean
and variances. The standard deviations of these vertical slices are shown in Fig.
\ref{fig:2dhisto_deriv1_var}, which exhibit a slow decrease with system size.
Similarly, for the 2D histograms of the $\partial \bar{S}/\partial \ell$ and $L \partial^2 \bar{S}/\partial \ell^2$, we also looked at the distribution 
of vertical slices of the two dimensional histograms. The standard deviations of these vertical slices are shown in Fig.
\ref{fig:2dhisto_deriv2_var}, which exhibits a slow decrease with system size. Calculating the third and
fourth cumulants also suggests that these distributions become more and more Gaussian with increasing system sizes.

\bibliography{mbl_entanglement_automatic,mbl_entanglement_manual,SSA_automatic}

\begin{thebibliography}{57}%
\makeatletter
\providecommand \@ifxundefined [1]{%
 \@ifx{#1\undefined}
}%
\providecommand \@ifnum [1]{%
 \ifnum #1\expandafter \@firstoftwo
 \else \expandafter \@secondoftwo
 \fi
}%
\providecommand \@ifx [1]{%
 \ifx #1\expandafter \@firstoftwo
 \else \expandafter \@secondoftwo
 \fi
}%
\providecommand \natexlab [1]{#1}%
\providecommand \enquote  [1]{``#1''}%
\providecommand \bibnamefont  [1]{#1}%
\providecommand \bibfnamefont [1]{#1}%
\providecommand \citenamefont [1]{#1}%
\providecommand \href@noop [0]{\@secondoftwo}%
\providecommand \href [0]{\begingroup \@sanitize@url \@href}%
\providecommand \@href[1]{\@@startlink{#1}\@@href}%
\providecommand \@@href[1]{\endgroup#1\@@endlink}%
\providecommand \@sanitize@url [0]{\catcode `\\12\catcode `\$12\catcode
  `\&12\catcode `\#12\catcode `\^12\catcode `\_12\catcode `\%12\relax}%
\providecommand \@@startlink[1]{}%
\providecommand \@@endlink[0]{}%
\providecommand \url  [0]{\begingroup\@sanitize@url \@url }%
\providecommand \@url [1]{\endgroup\@href {#1}{\urlprefix }}%
\providecommand \urlprefix  [0]{URL }%
\providecommand \Eprint [0]{\href }%
\providecommand \doibase [0]{http://dx.doi.org/}%
\providecommand \selectlanguage [0]{\@gobble}%
\providecommand \bibinfo  [0]{\@secondoftwo}%
\providecommand \bibfield  [0]{\@secondoftwo}%
\providecommand \translation [1]{[#1]}%
\providecommand \BibitemOpen [0]{}%
\providecommand \bibitemStop [0]{}%
\providecommand \bibitemNoStop [0]{.\EOS\space}%
\providecommand \EOS [0]{\spacefactor3000\relax}%
\providecommand \BibitemShut  [1]{\csname bibitem#1\endcsname}%
\let\auto@bib@innerbib\@empty
\bibitem [{\citenamefont {Basko}\ \emph {et~al.}(2006)\citenamefont {Basko},
  \citenamefont {Aleiner},\ and\ \citenamefont
  {Altshuler}}]{basko_metalinsulator_2006}%
  \BibitemOpen
  \bibfield  {author} {\bibinfo {author} {\bibfnamefont {D.~M.}\ \bibnamefont
  {Basko}}, \bibinfo {author} {\bibfnamefont {I.~L.}\ \bibnamefont {Aleiner}},
  \ and\ \bibinfo {author} {\bibfnamefont {B.~L.}\ \bibnamefont {Altshuler}},\
  }\bibfield  {title} {\enquote {\bibinfo {title} {Metal{\textendash}insulator
  transition in a weakly interacting many-electron system with localized
  single-particle states},}\ }\href {\doibase 10.1016/j.aop.2005.11.014}
  {\bibfield  {journal} {\bibinfo  {journal} {Annals of Physics}\ }\textbf
  {\bibinfo {volume} {321}},\ \bibinfo {pages} {1126--1205} (\bibinfo {year}
  {2006})}\BibitemShut {NoStop}%
\bibitem [{\citenamefont {Gornyi}\ \emph {et~al.}(2005)\citenamefont {Gornyi},
  \citenamefont {Mirlin},\ and\ \citenamefont
  {Polyakov}}]{gornyi_interacting_2005}%
  \BibitemOpen
  \bibfield  {author} {\bibinfo {author} {\bibfnamefont {I.~V.}\ \bibnamefont
  {Gornyi}}, \bibinfo {author} {\bibfnamefont {A.~D.}\ \bibnamefont {Mirlin}},
  \ and\ \bibinfo {author} {\bibfnamefont {D.~G.}\ \bibnamefont {Polyakov}},\
  }\bibfield  {title} {\enquote {\bibinfo {title} {Interacting {Electrons} in
  {Disordered} {Wires}: {Anderson} {Localization} and {Low}-{T} {Transport}},}\
  }\href {\doibase 10.1103/PhysRevLett.95.206603} {\bibfield  {journal}
  {\bibinfo  {journal} {Phys. Rev. Lett.}\ }\textbf {\bibinfo {volume} {95}},\
  \bibinfo {pages} {206603} (\bibinfo {year} {2005})}\BibitemShut {NoStop}%
\bibitem [{\citenamefont {Anderson}(1958)}]{anderson_absence_1958}%
  \BibitemOpen
  \bibfield  {author} {\bibinfo {author} {\bibfnamefont {P.~W.}\ \bibnamefont
  {Anderson}},\ }\bibfield  {title} {\enquote {\bibinfo {title} {Absence of
  {Diffusion} in {Certain} {Random} {Lattices}},}\ }\href {\doibase
  10.1103/PhysRev.109.1492} {\bibfield  {journal} {\bibinfo  {journal} {Phys.
  Rev.}\ }\textbf {\bibinfo {volume} {109}},\ \bibinfo {pages} {1492--1505}
  (\bibinfo {year} {1958})}\BibitemShut {NoStop}%
\bibitem [{\citenamefont {Deutsch}(1991)}]{deutsch_quantum_1991}%
  \BibitemOpen
  \bibfield  {author} {\bibinfo {author} {\bibfnamefont {J.~M.}\ \bibnamefont
  {Deutsch}},\ }\bibfield  {title} {\enquote {\bibinfo {title} {Quantum
  statistical mechanics in a closed system},}\ }\href {\doibase
  10.1103/PhysRevA.43.2046} {\bibfield  {journal} {\bibinfo  {journal} {Phys.
  Rev. A}\ }\textbf {\bibinfo {volume} {43}},\ \bibinfo {pages} {2046--2049}
  (\bibinfo {year} {1991})}\BibitemShut {NoStop}%
\bibitem [{\citenamefont {Srednicki}(1994)}]{srednicki_chaos_1994}%
  \BibitemOpen
  \bibfield  {author} {\bibinfo {author} {\bibfnamefont {Mark}\ \bibnamefont
  {Srednicki}},\ }\bibfield  {title} {\enquote {\bibinfo {title} {Chaos and
  quantum thermalization},}\ }\href {\doibase 10.1103/PhysRevE.50.888}
  {\bibfield  {journal} {\bibinfo  {journal} {Phys. Rev. E}\ }\textbf {\bibinfo
  {volume} {50}},\ \bibinfo {pages} {888--901} (\bibinfo {year}
  {1994})}\BibitemShut {NoStop}%
\bibitem [{\citenamefont {Montambaux}\ \emph {et~al.}(1993)\citenamefont
  {Montambaux}, \citenamefont {Poilblanc}, \citenamefont {Bellissard},\ and\
  \citenamefont {Sire}}]{montambaux_quantum_1993}%
  \BibitemOpen
  \bibfield  {author} {\bibinfo {author} {\bibfnamefont {Gilles}\ \bibnamefont
  {Montambaux}}, \bibinfo {author} {\bibfnamefont {Didier}\ \bibnamefont
  {Poilblanc}}, \bibinfo {author} {\bibfnamefont {Jean}\ \bibnamefont
  {Bellissard}}, \ and\ \bibinfo {author} {\bibfnamefont {Cl{\'e}ment}\
  \bibnamefont {Sire}},\ }\bibfield  {title} {\enquote {\bibinfo {title}
  {Quantum chaos in spin-fermion models},}\ }\href {\doibase
  10.1103/PhysRevLett.70.497} {\bibfield  {journal} {\bibinfo  {journal} {Phys.
  Rev. Lett.}\ }\textbf {\bibinfo {volume} {70}},\ \bibinfo {pages} {497--500}
  (\bibinfo {year} {1993})}\BibitemShut {NoStop}%
\bibitem [{\citenamefont {Jacquod}\ and\ \citenamefont
  {Shepelyansky}(1997)}]{jacquod_emergence_1997}%
  \BibitemOpen
  \bibfield  {author} {\bibinfo {author} {\bibfnamefont {Ph.}\ \bibnamefont
  {Jacquod}}\ and\ \bibinfo {author} {\bibfnamefont {D.~L.}\ \bibnamefont
  {Shepelyansky}},\ }\bibfield  {title} {\enquote {\bibinfo {title} {Emergence
  of {Quantum} {Chaos} in {Finite} {Interacting} {Fermi} {Systems}},}\ }\href
  {\doibase 10.1103/PhysRevLett.79.1837} {\bibfield  {journal} {\bibinfo
  {journal} {Phys. Rev. Lett.}\ }\textbf {\bibinfo {volume} {79}},\ \bibinfo
  {pages} {1837--1840} (\bibinfo {year} {1997})}\BibitemShut {NoStop}%
\bibitem [{\citenamefont {Georgeot}\ and\ \citenamefont
  {Shepelyansky}(1998)}]{georgeot_integrability_1998}%
  \BibitemOpen
  \bibfield  {author} {\bibinfo {author} {\bibfnamefont {B.}~\bibnamefont
  {Georgeot}}\ and\ \bibinfo {author} {\bibfnamefont {D.~L.}\ \bibnamefont
  {Shepelyansky}},\ }\bibfield  {title} {\enquote {\bibinfo {title}
  {Integrability and {Quantum} {Chaos} in {Spin} {Glass} {Shards}},}\ }\href
  {\doibase 10.1103/PhysRevLett.81.5129} {\bibfield  {journal} {\bibinfo
  {journal} {Phys. Rev. Lett.}\ }\textbf {\bibinfo {volume} {81}},\ \bibinfo
  {pages} {5129--5132} (\bibinfo {year} {1998})}\BibitemShut {NoStop}%
\bibitem [{\citenamefont {Song}\ and\ \citenamefont
  {Shepelyansky}(2000)}]{song_low-energy_2000}%
  \BibitemOpen
  \bibfield  {author} {\bibinfo {author} {\bibfnamefont {Pil~Hun}\ \bibnamefont
  {Song}}\ and\ \bibinfo {author} {\bibfnamefont {Dima~L.}\ \bibnamefont
  {Shepelyansky}},\ }\bibfield  {title} {\enquote {\bibinfo {title} {Low-energy
  transition in spectral statistics of two-dimensional interacting fermions},}\
  }\href {\doibase 10.1103/PhysRevB.61.15546} {\bibfield  {journal} {\bibinfo
  {journal} {Phys. Rev. B}\ }\textbf {\bibinfo {volume} {61}},\ \bibinfo
  {pages} {15546--15549} (\bibinfo {year} {2000})}\BibitemShut {NoStop}%
\bibitem [{\citenamefont {Oganesyan}\ and\ \citenamefont
  {Huse}(2007)}]{oganesyan_localization_2007}%
  \BibitemOpen
  \bibfield  {author} {\bibinfo {author} {\bibfnamefont {Vadim}\ \bibnamefont
  {Oganesyan}}\ and\ \bibinfo {author} {\bibfnamefont {David~A.}\ \bibnamefont
  {Huse}},\ }\bibfield  {title} {\enquote {\bibinfo {title} {Localization of
  interacting fermions at high temperature},}\ }\href {\doibase
  10.1103/PhysRevB.75.155111} {\bibfield  {journal} {\bibinfo  {journal} {Phys.
  Rev. B}\ }\textbf {\bibinfo {volume} {75}},\ \bibinfo {pages} {155111}
  (\bibinfo {year} {2007})}\BibitemShut {NoStop}%
\bibitem [{\citenamefont {Pal}\ and\ \citenamefont
  {Huse}(2010)}]{pal_many-body_2010}%
  \BibitemOpen
  \bibfield  {author} {\bibinfo {author} {\bibfnamefont {Arijeet}\ \bibnamefont
  {Pal}}\ and\ \bibinfo {author} {\bibfnamefont {David~A.}\ \bibnamefont
  {Huse}},\ }\bibfield  {title} {\enquote {\bibinfo {title} {Many-body
  localization phase transition},}\ }\href {\doibase
  10.1103/PhysRevB.82.174411} {\bibfield  {journal} {\bibinfo  {journal} {Phys.
  Rev. B}\ }\textbf {\bibinfo {volume} {82}},\ \bibinfo {pages} {174411}
  (\bibinfo {year} {2010})}\BibitemShut {NoStop}%
\bibitem [{\citenamefont {Luitz}\ \emph {et~al.}(2015)\citenamefont {Luitz},
  \citenamefont {Laflorencie},\ and\ \citenamefont
  {Alet}}]{luitz_many-body_2015}%
  \BibitemOpen
  \bibfield  {author} {\bibinfo {author} {\bibfnamefont {David~J.}\
  \bibnamefont {Luitz}}, \bibinfo {author} {\bibfnamefont {Nicolas}\
  \bibnamefont {Laflorencie}}, \ and\ \bibinfo {author} {\bibfnamefont
  {Fabien}\ \bibnamefont {Alet}},\ }\bibfield  {title} {\enquote {\bibinfo
  {title} {Many-body localization edge in the random-field {Heisenberg}
  chain},}\ }\href {\doibase 10.1103/PhysRevB.91.081103} {\bibfield  {journal}
  {\bibinfo  {journal} {Phys. Rev. B}\ }\textbf {\bibinfo {volume} {91}},\
  \bibinfo {pages} {081103} (\bibinfo {year} {2015})}\BibitemShut {NoStop}%
\bibitem [{\citenamefont {Serbyn}\ and\ \citenamefont
  {Moore}(2016)}]{serbyn_spectral_2016}%
  \BibitemOpen
  \bibfield  {author} {\bibinfo {author} {\bibfnamefont {Maksym}\ \bibnamefont
  {Serbyn}}\ and\ \bibinfo {author} {\bibfnamefont {Joel~E.}\ \bibnamefont
  {Moore}},\ }\bibfield  {title} {\enquote {\bibinfo {title} {Spectral
  statistics across the many-body localization transition},}\ }\href {\doibase
  10.1103/PhysRevB.93.041424} {\bibfield  {journal} {\bibinfo  {journal} {Phys.
  Rev. B}\ }\textbf {\bibinfo {volume} {93}},\ \bibinfo {pages} {041424}
  (\bibinfo {year} {2016})}\BibitemShut {NoStop}%
\bibitem [{\citenamefont {Monthus}(2016{\natexlab{a}})}]{monthus_level_2016}%
  \BibitemOpen
  \bibfield  {author} {\bibinfo {author} {\bibfnamefont {Cecile}\ \bibnamefont
  {Monthus}},\ }\bibfield  {title} {\enquote {\bibinfo {title} {Level repulsion
  exponent beta for {Many}-{Body} {Localization} {Transitions} and for
  {Anderson} {Localization} {Transitions} via {Dyson} {Brownian} {Motion}},}\
  }\href {\doibase 10.1088/1742-5468/2016/03/033113} {\bibfield  {journal}
  {\bibinfo  {journal} {Journal of Statistical Mechanics: Theory and
  Experiment}\ }\textbf {\bibinfo {volume} {2016}},\ \bibinfo {pages} {033113}
  (\bibinfo {year} {2016}{\natexlab{a}})},\ \bibinfo {note} {arXiv:
  1510.08322}\BibitemShut {NoStop}%
\bibitem [{\citenamefont {Serbyn}\ \emph {et~al.}(2013)\citenamefont {Serbyn},
  \citenamefont {Papi{\'c}},\ and\ \citenamefont {Abanin}}]{serbyn_local_2013}%
  \BibitemOpen
  \bibfield  {author} {\bibinfo {author} {\bibfnamefont {Maksym}\ \bibnamefont
  {Serbyn}}, \bibinfo {author} {\bibfnamefont {Z.}~\bibnamefont {Papi{\'c}}}, \
  and\ \bibinfo {author} {\bibfnamefont {Dmitry~A.}\ \bibnamefont {Abanin}},\
  }\bibfield  {title} {\enquote {\bibinfo {title} {Local {Conservation} {Laws}
  and the {Structure} of the {Many}-{Body} {Localized} {States}},}\ }\href
  {\doibase 10.1103/PhysRevLett.111.127201} {\bibfield  {journal} {\bibinfo
  {journal} {Phys. Rev. Lett.}\ }\textbf {\bibinfo {volume} {111}},\ \bibinfo
  {pages} {127201} (\bibinfo {year} {2013})}\BibitemShut {NoStop}%
\bibitem [{\citenamefont {Imbrie}(2014)}]{imbrie_many-body_2014}%
  \BibitemOpen
  \bibfield  {author} {\bibinfo {author} {\bibfnamefont {John~Z.}\ \bibnamefont
  {Imbrie}},\ }\bibfield  {title} {\enquote {\bibinfo {title} {On {Many}-{Body}
  {Localization} for {Quantum} {Spin} {Chains}},}\ }\href
  {http://arxiv.org/abs/1403.7837} {\bibfield  {journal} {\bibinfo  {journal}
  {arXiv:1403.7837 [cond-mat, physics:math-ph]}\ } (\bibinfo {year} {2014})},\
  \bibinfo {note} {arXiv: 1403.7837}\BibitemShut {NoStop}%
\bibitem [{\citenamefont {Bauer}\ and\ \citenamefont
  {Nayak}(2013)}]{bauer_area_2013}%
  \BibitemOpen
  \bibfield  {author} {\bibinfo {author} {\bibfnamefont {Bela}\ \bibnamefont
  {Bauer}}\ and\ \bibinfo {author} {\bibfnamefont {Chetan}\ \bibnamefont
  {Nayak}},\ }\bibfield  {title} {\enquote {\bibinfo {title} {Area laws in a
  many-body localized state and its implications for topological order},}\
  }\href {\doibase 10.1088/1742-5468/2013/09/P09005} {\bibfield  {journal}
  {\bibinfo  {journal} {J. Stat. Mech.}\ }\textbf {\bibinfo {volume} {2013}},\
  \bibinfo {pages} {P09005} (\bibinfo {year} {2013})}\BibitemShut {NoStop}%
\bibitem [{\citenamefont {Luitz}(2016)}]{luitz_long_2016}%
  \BibitemOpen
  \bibfield  {author} {\bibinfo {author} {\bibfnamefont {David~J.}\
  \bibnamefont {Luitz}},\ }\bibfield  {title} {\enquote {\bibinfo {title} {Long
  tail distributions near the many-body localization transition},}\ }\href
  {\doibase 10.1103/PhysRevB.93.134201} {\bibfield  {journal} {\bibinfo
  {journal} {Phys. Rev. B}\ }\textbf {\bibinfo {volume} {93}},\ \bibinfo
  {pages} {134201} (\bibinfo {year} {2016})}\BibitemShut {NoStop}%
\bibitem [{\citenamefont {Yu}\ \emph {et~al.}(2015)\citenamefont {Yu},
  \citenamefont {Pekker},\ and\ \citenamefont {Clark}}]{yu_finding_2015}%
  \BibitemOpen
  \bibfield  {author} {\bibinfo {author} {\bibfnamefont {Xiongjie}\
  \bibnamefont {Yu}}, \bibinfo {author} {\bibfnamefont {David}\ \bibnamefont
  {Pekker}}, \ and\ \bibinfo {author} {\bibfnamefont {Bryan~K.}\ \bibnamefont
  {Clark}},\ }\bibfield  {title} {\enquote {\bibinfo {title} {Finding matrix
  product state representations of highly-excited eigenstates of many-body
  localized {Hamiltonians}},}\ }\href {http://arxiv.org/abs/1509.01244}
  {\bibfield  {journal} {\bibinfo  {journal} {arXiv:1509.01244 [cond-mat]}\ }
  (\bibinfo {year} {2015})},\ \bibinfo {note} {arXiv: 1509.01244}\BibitemShut
  {NoStop}%
\bibitem [{\citenamefont {Khemani}\ \emph {et~al.}(2015)\citenamefont
  {Khemani}, \citenamefont {Pollmann},\ and\ \citenamefont
  {Sondhi}}]{khemani_obtaining_2015}%
  \BibitemOpen
  \bibfield  {author} {\bibinfo {author} {\bibfnamefont {Vedika}\ \bibnamefont
  {Khemani}}, \bibinfo {author} {\bibfnamefont {Frank}\ \bibnamefont
  {Pollmann}}, \ and\ \bibinfo {author} {\bibfnamefont {S.~L.}\ \bibnamefont
  {Sondhi}},\ }\bibfield  {title} {\enquote {\bibinfo {title} {Obtaining
  highly-excited eigenstates of many-body localized {Hamiltonians} by the
  density matrix renormalization group},}\ }\href
  {http://arxiv.org/abs/1509.00483} {\bibfield  {journal} {\bibinfo  {journal}
  {arXiv:1509.00483 [cond-mat]}\ } (\bibinfo {year} {2015})},\ \bibinfo {note}
  {arXiv: 1509.00483}\BibitemShut {NoStop}%
\bibitem [{\citenamefont {Lim}\ and\ \citenamefont
  {Sheng}(2015)}]{lim_nature_2015}%
  \BibitemOpen
  \bibfield  {author} {\bibinfo {author} {\bibfnamefont {S.~P.}\ \bibnamefont
  {Lim}}\ and\ \bibinfo {author} {\bibfnamefont {D.~N.}\ \bibnamefont
  {Sheng}},\ }\bibfield  {title} {\enquote {\bibinfo {title} {Nature of
  {Many}-{Body} {Localization} and {Transitions} by {Density} {Matrix}
  {Renormaliztion} {Group} and {Exact} {Diagonalization} {Studies}},}\ }\href
  {http://arxiv.org/abs/1510.08145} {\bibfield  {journal} {\bibinfo  {journal}
  {arXiv:1510.08145 [cond-mat]}\ } (\bibinfo {year} {2015})},\ \bibinfo {note}
  {arXiv: 1510.08145}\BibitemShut {NoStop}%
\bibitem [{\citenamefont {Garrison}\ and\ \citenamefont
  {Grover}(2015)}]{garrison_does_2015}%
  \BibitemOpen
  \bibfield  {author} {\bibinfo {author} {\bibfnamefont {James~R.}\
  \bibnamefont {Garrison}}\ and\ \bibinfo {author} {\bibfnamefont {Tarun}\
  \bibnamefont {Grover}},\ }\bibfield  {title} {\enquote {\bibinfo {title}
  {Does a single eigenstate encode the full {Hamiltonian}?}}\ }\href
  {http://arxiv.org/abs/1503.00729} {\bibfield  {journal} {\bibinfo  {journal}
  {arXiv:1503.00729 [cond-mat, physics:hep-th, physics:quant-ph]}\ } (\bibinfo
  {year} {2015})},\ \bibinfo {note} {arXiv: 1503.00729}\BibitemShut {NoStop}%
\bibitem [{\citenamefont {Grover}(2014)}]{grover_certain_2014}%
  \BibitemOpen
  \bibfield  {author} {\bibinfo {author} {\bibfnamefont {Tarun}\ \bibnamefont
  {Grover}},\ }\bibfield  {title} {\enquote {\bibinfo {title} {Certain
  {General} {Constraints} on the {Many}-{Body} {Localization} {Transition}},}\
  }\href {http://arxiv.org/abs/1405.1471} {\bibfield  {journal} {\bibinfo
  {journal} {arXiv:1405.1471 [cond-mat, physics:quant-ph]}\ } (\bibinfo {year}
  {2014})},\ \bibinfo {note} {arXiv: 1405.1471}\BibitemShut {NoStop}%
\bibitem [{\citenamefont {Vosk}\ \emph {et~al.}(2015)\citenamefont {Vosk},
  \citenamefont {Huse},\ and\ \citenamefont {Altman}}]{vosk_theory_2015}%
  \BibitemOpen
  \bibfield  {author} {\bibinfo {author} {\bibfnamefont {Ronen}\ \bibnamefont
  {Vosk}}, \bibinfo {author} {\bibfnamefont {David~A.}\ \bibnamefont {Huse}}, \
  and\ \bibinfo {author} {\bibfnamefont {Ehud}\ \bibnamefont {Altman}},\
  }\bibfield  {title} {\enquote {\bibinfo {title} {Theory of the {Many}-{Body}
  {Localization} {Transition} in {One}-{Dimensional} {Systems}},}\ }\href
  {\doibase 10.1103/PhysRevX.5.031032} {\bibfield  {journal} {\bibinfo
  {journal} {Phys. Rev. X}\ }\textbf {\bibinfo {volume} {5}},\ \bibinfo {pages}
  {031032} (\bibinfo {year} {2015})}\BibitemShut {NoStop}%
\bibitem [{\citenamefont
  {Monthus}(2016{\natexlab{b}})}]{monthus_many-body-localization_2016}%
  \BibitemOpen
  \bibfield  {author} {\bibinfo {author} {\bibfnamefont {Cecile}\ \bibnamefont
  {Monthus}},\ }\bibfield  {title} {\enquote {\bibinfo {title}
  {Many-{Body}-{Localization} {Transition} : strong multifractality spectrum
  for matrix elements of local operators},}\ }\href
  {http://arxiv.org/abs/1603.04701} {\bibfield  {journal} {\bibinfo  {journal}
  {arXiv:1603.04701 [cond-mat]}\ } (\bibinfo {year} {2016}{\natexlab{b}})},\
  \bibinfo {note} {arXiv: 1603.04701}\BibitemShut {NoStop}%
\bibitem [{\citenamefont {Monthus}(2016{\natexlab{c}})}]{monthus_many_2016}%
  \BibitemOpen
  \bibfield  {author} {\bibinfo {author} {\bibfnamefont {Cecile}\ \bibnamefont
  {Monthus}},\ }\bibfield  {title} {\enquote {\bibinfo {title} {Many {Body}
  {Localization} {Transition} in the strong disorder limit : entanglement
  entropy from the statistics of rare extensive resonances},}\ }\href {\doibase
  10.3390/e18040122} {\bibfield  {journal} {\bibinfo  {journal} {Entropy}\
  }\textbf {\bibinfo {volume} {18}},\ \bibinfo {pages} {122} (\bibinfo {year}
  {2016}{\natexlab{c}})},\ \bibinfo {note} {arXiv: 1510.03711}\BibitemShut
  {NoStop}%
\bibitem [{\citenamefont {Luitz}\ and\ \citenamefont
  {Bar~Lev}(2016)}]{luitz_anomalous_2016}%
  \BibitemOpen
  \bibfield  {author} {\bibinfo {author} {\bibfnamefont {David~J.}\
  \bibnamefont {Luitz}}\ and\ \bibinfo {author} {\bibfnamefont {Yevgeny}\
  \bibnamefont {Bar~Lev}},\ }\bibfield  {title} {\enquote {\bibinfo {title}
  {Anomalous thermalization in ergodic phases},}\ }\href@noop {} {\  (\bibinfo
  {year} {2016})},\ \bibinfo {note} {unpublished}\BibitemShut {NoStop}%
\bibitem [{Note1()}]{Note1}%
  \BibitemOpen
  \bibinfo {note} {With R\'enyi index 1}\BibitemShut {NoStop}%
\bibitem [{\citenamefont {Lieb}\ and\ \citenamefont
  {Ruskai}(1973)}]{lieb_proof_1973}%
  \BibitemOpen
  \bibfield  {author} {\bibinfo {author} {\bibfnamefont {Elliott~H.}\
  \bibnamefont {Lieb}}\ and\ \bibinfo {author} {\bibfnamefont {Mary~Beth}\
  \bibnamefont {Ruskai}},\ }\bibfield  {title} {\enquote {\bibinfo {title}
  {Proof of the strong subadditivity of quantum-mechanical entropy},}\ }\href
  {\doibase 10.1063/1.1666274} {\bibfield  {journal} {\bibinfo  {journal}
  {Journal of Mathematical Physics}\ }\textbf {\bibinfo {volume} {14}},\
  \bibinfo {pages} {1938--1941} (\bibinfo {year} {1973})}\BibitemShut {NoStop}%
\bibitem [{\citenamefont {Lieb}(1975)}]{lieb_convexity_1975}%
  \BibitemOpen
  \bibfield  {author} {\bibinfo {author} {\bibfnamefont {Elliott~H.}\
  \bibnamefont {Lieb}},\ }\bibfield  {title} {\enquote {\bibinfo {title} {Some
  convexity and subadditivity properties of entropy},}\ }\href
  {http://projecteuclid.org/euclid.bams/1183536224} {\bibfield  {journal}
  {\bibinfo  {journal} {Bulletin of the American Mathematical Society}\
  }\textbf {\bibinfo {volume} {81}},\ \bibinfo {pages} {1--13} (\bibinfo {year}
  {1975})}\BibitemShut {NoStop}%
\bibitem [{\citenamefont {Lindblad}(1975)}]{lindblad_completely_1975}%
  \BibitemOpen
  \bibfield  {author} {\bibinfo {author} {\bibfnamefont {G{\"o}ran}\
  \bibnamefont {Lindblad}},\ }\bibfield  {title} {\enquote {\bibinfo {title}
  {Completely positive maps and entropy inequalities},}\ }\href {\doibase
  10.1007/BF01609396} {\bibfield  {journal} {\bibinfo  {journal}
  {Communications in Mathematical Physics}\ }\textbf {\bibinfo {volume} {40}},\
  \bibinfo {pages} {147--151} (\bibinfo {year} {1975})}\BibitemShut {NoStop}%
\bibitem [{\citenamefont {Ruskai}(2002)}]{ruskai_inequalities_2002}%
  \BibitemOpen
  \bibfield  {author} {\bibinfo {author} {\bibfnamefont {Mary~Beth}\
  \bibnamefont {Ruskai}},\ }\bibfield  {title} {\enquote {\bibinfo {title}
  {Inequalities for quantum entropy: {A} review with conditions for
  equality},}\ }\href {\doibase 10.1063/1.1497701} {\bibfield  {journal}
  {\bibinfo  {journal} {Journal of Mathematical Physics}\ }\textbf {\bibinfo
  {volume} {43}},\ \bibinfo {pages} {4358--4375} (\bibinfo {year}
  {2002})}\BibitemShut {NoStop}%
\bibitem [{\citenamefont {{\v Z}nidari{\v c}}\ \emph
  {et~al.}(2008)\citenamefont {{\v Z}nidari{\v c}}, \citenamefont {Prosen},\
  and\ \citenamefont {Prelov{\v s}ek}}]{znidaric_many-body_2008}%
  \BibitemOpen
  \bibfield  {author} {\bibinfo {author} {\bibfnamefont {Marko}\ \bibnamefont
  {{\v Z}nidari{\v c}}}, \bibinfo {author} {\bibfnamefont {Toma{\v z}}\
  \bibnamefont {Prosen}}, \ and\ \bibinfo {author} {\bibfnamefont {Peter}\
  \bibnamefont {Prelov{\v s}ek}},\ }\bibfield  {title} {\enquote {\bibinfo
  {title} {Many-body localization in the {Heisenberg} {XXZ} magnet in a random
  field},}\ }\href {\doibase 10.1103/PhysRevB.77.064426} {\bibfield  {journal}
  {\bibinfo  {journal} {Phys. Rev. B}\ }\textbf {\bibinfo {volume} {77}},\
  \bibinfo {pages} {064426} (\bibinfo {year} {2008})}\BibitemShut {NoStop}%
\bibitem [{\citenamefont {Luca}\ and\ \citenamefont
  {Scardicchio}(2013)}]{luca_ergodicity_2013}%
  \BibitemOpen
  \bibfield  {author} {\bibinfo {author} {\bibfnamefont {A.~De}\ \bibnamefont
  {Luca}}\ and\ \bibinfo {author} {\bibfnamefont {A.}~\bibnamefont
  {Scardicchio}},\ }\bibfield  {title} {\enquote {\bibinfo {title} {Ergodicity
  breaking in a model showing many-body localization},}\ }\href {\doibase
  10.1209/0295-5075/101/37003} {\bibfield  {journal} {\bibinfo  {journal}
  {EPL}\ }\textbf {\bibinfo {volume} {101}},\ \bibinfo {pages} {37003}
  (\bibinfo {year} {2013})}\BibitemShut {NoStop}%
\bibitem [{\citenamefont {Pekker}\ and\ \citenamefont
  {Clark}(2014)}]{pekker_encoding_2014}%
  \BibitemOpen
  \bibfield  {author} {\bibinfo {author} {\bibfnamefont {David}\ \bibnamefont
  {Pekker}}\ and\ \bibinfo {author} {\bibfnamefont {Bryan~K.}\ \bibnamefont
  {Clark}},\ }\bibfield  {title} {\enquote {\bibinfo {title} {Encoding the
  structure of many-body localization with matrix product operators},}\ }\href
  {http://arxiv.org/abs/1410.2224} {\bibfield  {journal} {\bibinfo  {journal}
  {arXiv:1410.2224 [cond-mat]}\ } (\bibinfo {year} {2014})},\ \bibinfo {note}
  {arXiv: 1410.2224}\BibitemShut {NoStop}%
\bibitem [{\citenamefont {Serbyn}\ \emph {et~al.}(2015)\citenamefont {Serbyn},
  \citenamefont {Papi{\'c}},\ and\ \citenamefont
  {Abanin}}]{serbyn_criterion_2015}%
  \BibitemOpen
  \bibfield  {author} {\bibinfo {author} {\bibfnamefont {Maksym}\ \bibnamefont
  {Serbyn}}, \bibinfo {author} {\bibfnamefont {Z.}~\bibnamefont {Papi{\'c}}}, \
  and\ \bibinfo {author} {\bibfnamefont {Dmitry~A.}\ \bibnamefont {Abanin}},\
  }\bibfield  {title} {\enquote {\bibinfo {title} {Criterion for {Many}-{Body}
  {Localization}-{Delocalization} {Phase} {Transition}},}\ }\href {\doibase
  10.1103/PhysRevX.5.041047} {\bibfield  {journal} {\bibinfo  {journal} {Phys.
  Rev. X}\ }\textbf {\bibinfo {volume} {5}},\ \bibinfo {pages} {041047}
  (\bibinfo {year} {2015})}\BibitemShut {NoStop}%
\bibitem [{\citenamefont {Bera}\ \emph {et~al.}(2015)\citenamefont {Bera},
  \citenamefont {Schomerus}, \citenamefont {Heidrich-Meisner},\ and\
  \citenamefont {Bardarson}}]{bera_many-body_2015}%
  \BibitemOpen
  \bibfield  {author} {\bibinfo {author} {\bibfnamefont {Soumya}\ \bibnamefont
  {Bera}}, \bibinfo {author} {\bibfnamefont {Henning}\ \bibnamefont
  {Schomerus}}, \bibinfo {author} {\bibfnamefont {Fabian}\ \bibnamefont
  {Heidrich-Meisner}}, \ and\ \bibinfo {author} {\bibfnamefont {Jens~H.}\
  \bibnamefont {Bardarson}},\ }\bibfield  {title} {\enquote {\bibinfo {title}
  {Many-{Body} {Localization} {Characterized} from a {One}-{Particle}
  {Perspective}},}\ }\href {\doibase 10.1103/PhysRevLett.115.046603} {\bibfield
   {journal} {\bibinfo  {journal} {Phys. Rev. Lett.}\ }\textbf {\bibinfo
  {volume} {115}},\ \bibinfo {pages} {046603} (\bibinfo {year}
  {2015})}\BibitemShut {NoStop}%
\bibitem [{\citenamefont {Luitz}\ \emph {et~al.}(2016)\citenamefont {Luitz},
  \citenamefont {Laflorencie},\ and\ \citenamefont
  {Alet}}]{luitz_extended_2016}%
  \BibitemOpen
  \bibfield  {author} {\bibinfo {author} {\bibfnamefont {David~J.}\
  \bibnamefont {Luitz}}, \bibinfo {author} {\bibfnamefont {Nicolas}\
  \bibnamefont {Laflorencie}}, \ and\ \bibinfo {author} {\bibfnamefont
  {Fabien}\ \bibnamefont {Alet}},\ }\bibfield  {title} {\enquote {\bibinfo
  {title} {Extended slow dynamical regime close to the many-body localization
  transition},}\ }\href {\doibase 10.1103/PhysRevB.93.060201} {\bibfield
  {journal} {\bibinfo  {journal} {Phys. Rev. B}\ }\textbf {\bibinfo {volume}
  {93}},\ \bibinfo {pages} {060201} (\bibinfo {year} {2016})}\BibitemShut
  {NoStop}%
\bibitem [{\citenamefont {Singh}\ \emph {et~al.}(2016)\citenamefont {Singh},
  \citenamefont {Bardarson},\ and\ \citenamefont
  {Pollmann}}]{singh_signatures_2016}%
  \BibitemOpen
  \bibfield  {author} {\bibinfo {author} {\bibfnamefont {Rajeev}\ \bibnamefont
  {Singh}}, \bibinfo {author} {\bibfnamefont {Jens~H.}\ \bibnamefont
  {Bardarson}}, \ and\ \bibinfo {author} {\bibfnamefont {Frank}\ \bibnamefont
  {Pollmann}},\ }\bibfield  {title} {\enquote {\bibinfo {title} {Signatures of
  the many-body localization transition in the dynamics of entanglement and
  bipartite fluctuations},}\ }\href {\doibase 10.1088/1367-2630/18/2/023046}
  {\bibfield  {journal} {\bibinfo  {journal} {New J. Phys.}\ }\textbf {\bibinfo
  {volume} {18}},\ \bibinfo {pages} {023046} (\bibinfo {year}
  {2016})}\BibitemShut {NoStop}%
\bibitem [{\citenamefont {Pollmann}\ \emph {et~al.}(2015)\citenamefont
  {Pollmann}, \citenamefont {Khemani}, \citenamefont {Cirac},\ and\
  \citenamefont {Sondhi}}]{pollmann_efficient_2015}%
  \BibitemOpen
  \bibfield  {author} {\bibinfo {author} {\bibfnamefont {Frank}\ \bibnamefont
  {Pollmann}}, \bibinfo {author} {\bibfnamefont {Vedika}\ \bibnamefont
  {Khemani}}, \bibinfo {author} {\bibfnamefont {J.~Ignacio}\ \bibnamefont
  {Cirac}}, \ and\ \bibinfo {author} {\bibfnamefont {S.~L.}\ \bibnamefont
  {Sondhi}},\ }\bibfield  {title} {\enquote {\bibinfo {title} {Efficient
  variational diagonalization of fully many-body localized {Hamiltonians}},}\
  }\href {http://arxiv.org/abs/1506.07179} {\bibfield  {journal} {\bibinfo
  {journal} {arXiv:1506.07179 [cond-mat]}\ } (\bibinfo {year} {2015})},\
  \bibinfo {note} {arXiv: 1506.07179}\BibitemShut {NoStop}%
\bibitem [{\citenamefont {Bera}\ and\ \citenamefont
  {Lakshminarayan}(2016)}]{bera_local_2016}%
  \BibitemOpen
  \bibfield  {author} {\bibinfo {author} {\bibfnamefont {Soumya}\ \bibnamefont
  {Bera}}\ and\ \bibinfo {author} {\bibfnamefont {Arul}\ \bibnamefont
  {Lakshminarayan}},\ }\bibfield  {title} {\enquote {\bibinfo {title} {Local
  entanglement structure across a many-body localization transition},}\ }\href
  {\doibase 10.1103/PhysRevB.93.134204} {\bibfield  {journal} {\bibinfo
  {journal} {Phys. Rev. B}\ }\textbf {\bibinfo {volume} {93}},\ \bibinfo
  {pages} {134204} (\bibinfo {year} {2016})}\BibitemShut {NoStop}%
\bibitem [{\citenamefont {Devakul}\ and\ \citenamefont
  {Singh}(2015)}]{devakul_early_2015}%
  \BibitemOpen
  \bibfield  {author} {\bibinfo {author} {\bibfnamefont {Trithep}\ \bibnamefont
  {Devakul}}\ and\ \bibinfo {author} {\bibfnamefont {Rajiv R.~P.}\ \bibnamefont
  {Singh}},\ }\bibfield  {title} {\enquote {\bibinfo {title} {Early {Breakdown}
  of {Area}-{Law} {Entanglement} at the {Many}-{Body} {Delocalization}
  {Transition}},}\ }\href {\doibase 10.1103/PhysRevLett.115.187201} {\bibfield
  {journal} {\bibinfo  {journal} {Phys. Rev. Lett.}\ }\textbf {\bibinfo
  {volume} {115}},\ \bibinfo {pages} {187201} (\bibinfo {year}
  {2015})}\BibitemShut {NoStop}%
\bibitem [{\citenamefont {Kj{\"a}ll}\ \emph {et~al.}(2014)\citenamefont
  {Kj{\"a}ll}, \citenamefont {Bardarson},\ and\ \citenamefont
  {Pollmann}}]{kjall_many-body_2014}%
  \BibitemOpen
  \bibfield  {author} {\bibinfo {author} {\bibfnamefont {Jonas~A.}\
  \bibnamefont {Kj{\"a}ll}}, \bibinfo {author} {\bibfnamefont {Jens~H.}\
  \bibnamefont {Bardarson}}, \ and\ \bibinfo {author} {\bibfnamefont {Frank}\
  \bibnamefont {Pollmann}},\ }\bibfield  {title} {\enquote {\bibinfo {title}
  {Many-{Body} {Localization} in a {Disordered} {Quantum} {Ising} {Chain}},}\
  }\href {\doibase 10.1103/PhysRevLett.113.107204} {\bibfield  {journal}
  {\bibinfo  {journal} {Phys. Rev. Lett.}\ }\textbf {\bibinfo {volume} {113}},\
  \bibinfo {pages} {107204} (\bibinfo {year} {2014})}\BibitemShut {NoStop}%
\bibitem [{\citenamefont {Chen}\ \emph {et~al.}(2015)\citenamefont {Chen},
  \citenamefont {Yu}, \citenamefont {Cho}, \citenamefont {Clark},\ and\
  \citenamefont {Fradkin}}]{chen_many-body_2015}%
  \BibitemOpen
  \bibfield  {author} {\bibinfo {author} {\bibfnamefont {Xiao}\ \bibnamefont
  {Chen}}, \bibinfo {author} {\bibfnamefont {Xiongjie}\ \bibnamefont {Yu}},
  \bibinfo {author} {\bibfnamefont {Gil~Young}\ \bibnamefont {Cho}}, \bibinfo
  {author} {\bibfnamefont {Bryan~K.}\ \bibnamefont {Clark}}, \ and\ \bibinfo
  {author} {\bibfnamefont {Eduardo}\ \bibnamefont {Fradkin}},\ }\bibfield
  {title} {\enquote {\bibinfo {title} {Many-body localization transition in
  {Rokhsar}-{Kivelson}-type wave functions},}\ }\href {\doibase
  10.1103/PhysRevB.92.214204} {\bibfield  {journal} {\bibinfo  {journal} {Phys.
  Rev. B}\ }\textbf {\bibinfo {volume} {92}},\ \bibinfo {pages} {214204}
  (\bibinfo {year} {2015})}\BibitemShut {NoStop}%
\bibitem [{Note2()}]{Note2}%
  \BibitemOpen
  \bibinfo {note} {We have tried a tentative extrapolation of the peak height
  to the thermodynamic limit using a polynomial ansatz as a function of $1/L$,
  pointing to a maximal variance of $\protect \qopname \relax o{ln}2/2$,
  although this extrapolation is not reliable, as the extrapolation typically
  overshoots the theoretical maximum due to finite size effects.}\BibitemShut
  {Stop}%
\bibitem [{\citenamefont {Potter}\ \emph {et~al.}(2015)\citenamefont {Potter},
  \citenamefont {Vasseur},\ and\ \citenamefont
  {Parameswaran}}]{potter_universal_2015}%
  \BibitemOpen
  \bibfield  {author} {\bibinfo {author} {\bibfnamefont {Andrew~C.}\
  \bibnamefont {Potter}}, \bibinfo {author} {\bibfnamefont {Romain}\
  \bibnamefont {Vasseur}}, \ and\ \bibinfo {author} {\bibfnamefont {S.~A.}\
  \bibnamefont {Parameswaran}},\ }\bibfield  {title} {\enquote {\bibinfo
  {title} {Universal {Properties} of {Many}-{Body} {Delocalization}
  {Transitions}},}\ }\href {\doibase 10.1103/PhysRevX.5.031033} {\bibfield
  {journal} {\bibinfo  {journal} {Phys. Rev. X}\ }\textbf {\bibinfo {volume}
  {5}},\ \bibinfo {pages} {031033} (\bibinfo {year} {2015})}\BibitemShut
  {NoStop}%
\bibitem [{\citenamefont {Bar~Lev}\ \emph {et~al.}(2015)\citenamefont
  {Bar~Lev}, \citenamefont {Cohen},\ and\ \citenamefont
  {Reichman}}]{bar_lev_absence_2015}%
  \BibitemOpen
  \bibfield  {author} {\bibinfo {author} {\bibfnamefont {Yevgeny}\ \bibnamefont
  {Bar~Lev}}, \bibinfo {author} {\bibfnamefont {Guy}\ \bibnamefont {Cohen}}, \
  and\ \bibinfo {author} {\bibfnamefont {David~R.}\ \bibnamefont {Reichman}},\
  }\bibfield  {title} {\enquote {\bibinfo {title} {Absence of {Diffusion} in an
  {Interacting} {System} of {Spinless} {Fermions} on a {One}-{Dimensional}
  {Disordered} {Lattice}},}\ }\href {\doibase 10.1103/PhysRevLett.114.100601}
  {\bibfield  {journal} {\bibinfo  {journal} {Phys. Rev. Lett.}\ }\textbf
  {\bibinfo {volume} {114}},\ \bibinfo {pages} {100601} (\bibinfo {year}
  {2015})}\BibitemShut {NoStop}%
\bibitem [{\citenamefont {Agarwal}\ \emph {et~al.}(2015)\citenamefont
  {Agarwal}, \citenamefont {Gopalakrishnan}, \citenamefont {Knap},
  \citenamefont {M{\"u}ller},\ and\ \citenamefont
  {Demler}}]{agarwal_anomalous_2015}%
  \BibitemOpen
  \bibfield  {author} {\bibinfo {author} {\bibfnamefont {Kartiek}\ \bibnamefont
  {Agarwal}}, \bibinfo {author} {\bibfnamefont {Sarang}\ \bibnamefont
  {Gopalakrishnan}}, \bibinfo {author} {\bibfnamefont {Michael}\ \bibnamefont
  {Knap}}, \bibinfo {author} {\bibfnamefont {Markus}\ \bibnamefont
  {M{\"u}ller}}, \ and\ \bibinfo {author} {\bibfnamefont {Eugene}\ \bibnamefont
  {Demler}},\ }\bibfield  {title} {\enquote {\bibinfo {title} {Anomalous
  {Diffusion} and {Griffiths} {Effects} {Near} the {Many}-{Body} {Localization}
  {Transition}},}\ }\href {\doibase 10.1103/PhysRevLett.114.160401} {\bibfield
  {journal} {\bibinfo  {journal} {Phys. Rev. Lett.}\ }\textbf {\bibinfo
  {volume} {114}},\ \bibinfo {pages} {160401} (\bibinfo {year}
  {2015})}\BibitemShut {NoStop}%
\bibitem [{\citenamefont {Varma}\ \emph {et~al.}(2015)\citenamefont {Varma},
  \citenamefont {Lerose}, \citenamefont {Pietracaprina}, \citenamefont
  {Goold},\ and\ \citenamefont {Scardicchio}}]{varma_energy_2015}%
  \BibitemOpen
  \bibfield  {author} {\bibinfo {author} {\bibfnamefont {Vipin~Kerala}\
  \bibnamefont {Varma}}, \bibinfo {author} {\bibfnamefont {Alessio}\
  \bibnamefont {Lerose}}, \bibinfo {author} {\bibfnamefont {Francesca}\
  \bibnamefont {Pietracaprina}}, \bibinfo {author} {\bibfnamefont {John}\
  \bibnamefont {Goold}}, \ and\ \bibinfo {author} {\bibfnamefont {Antonello}\
  \bibnamefont {Scardicchio}},\ }\bibfield  {title} {\enquote {\bibinfo {title}
  {Energy diffusion in the ergodic phase of a many body localizable spin
  chain},}\ }\href {http://arxiv.org/abs/1511.09144} {\bibfield  {journal}
  {\bibinfo  {journal} {arXiv:1511.09144 [cond-mat, physics:quant-ph]}\ }
  (\bibinfo {year} {2015})},\ \bibinfo {note} {arXiv: 1511.09144}\BibitemShut
  {NoStop}%
\bibitem [{\citenamefont {Gopalakrishnan}\ \emph {et~al.}(2016)\citenamefont
  {Gopalakrishnan}, \citenamefont {Agarwal}, \citenamefont {Demler},
  \citenamefont {Huse},\ and\ \citenamefont
  {Knap}}]{gopalakrishnan_griffiths_2016}%
  \BibitemOpen
  \bibfield  {author} {\bibinfo {author} {\bibfnamefont {Sarang}\ \bibnamefont
  {Gopalakrishnan}}, \bibinfo {author} {\bibfnamefont {Kartiek}\ \bibnamefont
  {Agarwal}}, \bibinfo {author} {\bibfnamefont {Eugene~A.}\ \bibnamefont
  {Demler}}, \bibinfo {author} {\bibfnamefont {David~A.}\ \bibnamefont {Huse}},
  \ and\ \bibinfo {author} {\bibfnamefont {Michael}\ \bibnamefont {Knap}},\
  }\bibfield  {title} {\enquote {\bibinfo {title} {Griffiths effects and slow
  dynamics in nearly many-body localized systems},}\ }\href {\doibase
  10.1103/PhysRevB.93.134206} {\bibfield  {journal} {\bibinfo  {journal} {Phys.
  Rev. B}\ }\textbf {\bibinfo {volume} {93}},\ \bibinfo {pages} {134206}
  (\bibinfo {year} {2016})}\BibitemShut {NoStop}%
\bibitem [{\citenamefont {Pekker}\ \emph {et~al.}(2014)\citenamefont {Pekker},
  \citenamefont {Refael}, \citenamefont {Altman}, \citenamefont {Demler},\ and\
  \citenamefont {Oganesyan}}]{pekker_hilbert-glass_2014}%
  \BibitemOpen
  \bibfield  {author} {\bibinfo {author} {\bibfnamefont {David}\ \bibnamefont
  {Pekker}}, \bibinfo {author} {\bibfnamefont {Gil}\ \bibnamefont {Refael}},
  \bibinfo {author} {\bibfnamefont {Ehud}\ \bibnamefont {Altman}}, \bibinfo
  {author} {\bibfnamefont {Eugene}\ \bibnamefont {Demler}}, \ and\ \bibinfo
  {author} {\bibfnamefont {Vadim}\ \bibnamefont {Oganesyan}},\ }\bibfield
  {title} {\enquote {\bibinfo {title} {Hilbert-{Glass} {Transition}: {New}
  {Universality} of {Temperature}-{Tuned} {Many}-{Body} {Dynamical} {Quantum}
  {Criticality}},}\ }\href {\doibase 10.1103/PhysRevX.4.011052} {\bibfield
  {journal} {\bibinfo  {journal} {Phys. Rev. X}\ }\textbf {\bibinfo {volume}
  {4}},\ \bibinfo {pages} {011052} (\bibinfo {year} {2014})}\BibitemShut
  {NoStop}%
\bibitem [{\citenamefont {Balay}\ \emph
  {et~al.}(2014{\natexlab{a}})\citenamefont {Balay}, \citenamefont {Abhyankar},
  \citenamefont {Adams}, \citenamefont {Brown}, \citenamefont {Brune},
  \citenamefont {Buschelman}, \citenamefont {Eijkhout}, \citenamefont {Gropp},
  \citenamefont {Kaushik}, \citenamefont {Knepley}, \citenamefont {McInnes},
  \citenamefont {Rupp}, \citenamefont {Smith},\ and\ \citenamefont
  {Zhang}}]{petsc-web-page}%
  \BibitemOpen
  \bibfield  {author} {\bibinfo {author} {\bibfnamefont {Satish}\ \bibnamefont
  {Balay}}, \bibinfo {author} {\bibfnamefont {Shrirang}\ \bibnamefont
  {Abhyankar}}, \bibinfo {author} {\bibfnamefont {Mark~F.}\ \bibnamefont
  {Adams}}, \bibinfo {author} {\bibfnamefont {Jed}\ \bibnamefont {Brown}},
  \bibinfo {author} {\bibfnamefont {Peter}\ \bibnamefont {Brune}}, \bibinfo
  {author} {\bibfnamefont {Kris}\ \bibnamefont {Buschelman}}, \bibinfo {author}
  {\bibfnamefont {Victor}\ \bibnamefont {Eijkhout}}, \bibinfo {author}
  {\bibfnamefont {William~D.}\ \bibnamefont {Gropp}}, \bibinfo {author}
  {\bibfnamefont {Dinesh}\ \bibnamefont {Kaushik}}, \bibinfo {author}
  {\bibfnamefont {Matthew~G.}\ \bibnamefont {Knepley}}, \bibinfo {author}
  {\bibfnamefont {Lois~Curfman}\ \bibnamefont {McInnes}}, \bibinfo {author}
  {\bibfnamefont {Karl}\ \bibnamefont {Rupp}}, \bibinfo {author} {\bibfnamefont
  {Barry~F.}\ \bibnamefont {Smith}}, \ and\ \bibinfo {author} {\bibfnamefont
  {Hong}\ \bibnamefont {Zhang}},\ }\href {http://www.mcs.anl.gov/petsc}
  {\enquote {\bibinfo {title} {{PETS}c {W}eb page},}\ }\bibinfo {howpublished}
  {\url{http://www.mcs.anl.gov/petsc}} (\bibinfo {year}
  {2014}{\natexlab{a}})\BibitemShut {NoStop}%
\bibitem [{\citenamefont {Balay}\ \emph
  {et~al.}(2014{\natexlab{b}})\citenamefont {Balay}, \citenamefont {Abhyankar},
  \citenamefont {Adams}, \citenamefont {Brown}, \citenamefont {Brune},
  \citenamefont {Buschelman}, \citenamefont {Eijkhout}, \citenamefont {Gropp},
  \citenamefont {Kaushik}, \citenamefont {Knepley}, \citenamefont {McInnes},
  \citenamefont {Rupp}, \citenamefont {Smith},\ and\ \citenamefont
  {Zhang}}]{petsc-user-ref}%
  \BibitemOpen
  \bibfield  {author} {\bibinfo {author} {\bibfnamefont {Satish}\ \bibnamefont
  {Balay}}, \bibinfo {author} {\bibfnamefont {Shrirang}\ \bibnamefont
  {Abhyankar}}, \bibinfo {author} {\bibfnamefont {Mark~F.}\ \bibnamefont
  {Adams}}, \bibinfo {author} {\bibfnamefont {Jed}\ \bibnamefont {Brown}},
  \bibinfo {author} {\bibfnamefont {Peter}\ \bibnamefont {Brune}}, \bibinfo
  {author} {\bibfnamefont {Kris}\ \bibnamefont {Buschelman}}, \bibinfo {author}
  {\bibfnamefont {Victor}\ \bibnamefont {Eijkhout}}, \bibinfo {author}
  {\bibfnamefont {William~D.}\ \bibnamefont {Gropp}}, \bibinfo {author}
  {\bibfnamefont {Dinesh}\ \bibnamefont {Kaushik}}, \bibinfo {author}
  {\bibfnamefont {Matthew~G.}\ \bibnamefont {Knepley}}, \bibinfo {author}
  {\bibfnamefont {Lois~Curfman}\ \bibnamefont {McInnes}}, \bibinfo {author}
  {\bibfnamefont {Karl}\ \bibnamefont {Rupp}}, \bibinfo {author} {\bibfnamefont
  {Barry~F.}\ \bibnamefont {Smith}}, \ and\ \bibinfo {author} {\bibfnamefont
  {Hong}\ \bibnamefont {Zhang}},\ }\href {http://www.mcs.anl.gov/petsc} {\emph
  {\bibinfo {title} {{PETS}c Users Manual}}},\ \bibinfo {type} {Tech. Rep.}\
  \bibinfo {number} {ANL-95/11 - Revision 3.5}\ (\bibinfo  {institution}
  {Argonne National Laboratory},\ \bibinfo {year} {2014})\BibitemShut {NoStop}%
\bibitem [{\citenamefont {Balay}\ \emph {et~al.}(1997)\citenamefont {Balay},
  \citenamefont {Gropp}, \citenamefont {McInnes},\ and\ \citenamefont
  {Smith}}]{petsc-efficient}%
  \BibitemOpen
  \bibfield  {author} {\bibinfo {author} {\bibfnamefont {Satish}\ \bibnamefont
  {Balay}}, \bibinfo {author} {\bibfnamefont {William~D.}\ \bibnamefont
  {Gropp}}, \bibinfo {author} {\bibfnamefont {Lois~Curfman}\ \bibnamefont
  {McInnes}}, \ and\ \bibinfo {author} {\bibfnamefont {Barry~F.}\ \bibnamefont
  {Smith}},\ }\bibfield  {title} {\enquote {\bibinfo {title} {Efficient
  management of parallelism in object oriented numerical software libraries},}\
  }in\ \href@noop {} {\emph {\bibinfo {booktitle} {Modern Software Tools in
  Scientific Computing}}},\ \bibinfo {editor} {edited by\ \bibinfo {editor}
  {\bibfnamefont {E.}~\bibnamefont {Arge}}, \bibinfo {editor} {\bibfnamefont
  {A.~M.}\ \bibnamefont {Bruaset}}, \ and\ \bibinfo {editor} {\bibfnamefont
  {H.~P.}\ \bibnamefont {Langtangen}}}\ (\bibinfo  {publisher}
  {Birkh{\"{a}}user Press},\ \bibinfo {year} {1997})\ pp.\ \bibinfo {pages}
  {163--202}\BibitemShut {NoStop}%
\bibitem [{\citenamefont {Hernandez}\ \emph {et~al.}(2005)\citenamefont
  {Hernandez}, \citenamefont {Roman},\ and\ \citenamefont
  {Vidal}}]{hernandez_slepc:_2005}%
  \BibitemOpen
  \bibfield  {author} {\bibinfo {author} {\bibfnamefont {Vicente}\ \bibnamefont
  {Hernandez}}, \bibinfo {author} {\bibfnamefont {Jose~E.}\ \bibnamefont
  {Roman}}, \ and\ \bibinfo {author} {\bibfnamefont {Vicente}\ \bibnamefont
  {Vidal}},\ }\bibfield  {title} {\enquote {\bibinfo {title} {{SLEPc}: {A}
  {Scalable} and {Flexible} {Toolkit} for the {Solution} of {Eigenvalue}
  {Problems}},}\ }\href {\doibase 10.1145/1089014.1089019} {\bibfield
  {journal} {\bibinfo  {journal} {ACM Trans. Math. Softw.}\ }\textbf {\bibinfo
  {volume} {31}},\ \bibinfo {pages} {351--362} (\bibinfo {year}
  {2005})}\BibitemShut {NoStop}%
\bibitem [{\citenamefont {Amestoy}\ \emph {et~al.}(2001)\citenamefont
  {Amestoy}, \citenamefont {Duff}, \citenamefont {Koster},\ and\ \citenamefont
  {L'Excellent}}]{MUMPS1}%
  \BibitemOpen
  \bibfield  {author} {\bibinfo {author} {\bibfnamefont {P.~R.}\ \bibnamefont
  {Amestoy}}, \bibinfo {author} {\bibfnamefont {I.~S.}\ \bibnamefont {Duff}},
  \bibinfo {author} {\bibfnamefont {J.}~\bibnamefont {Koster}}, \ and\ \bibinfo
  {author} {\bibfnamefont {J.-Y.}\ \bibnamefont {L'Excellent}},\ }\bibfield
  {title} {\enquote {\bibinfo {title} {A fully asynchronous multifrontal solver
  using distributed dynamic scheduling},}\ }\href@noop {} {\bibfield  {journal}
  {\bibinfo  {journal} {SIAM J. Matrix Anal. Appl.}\ }\textbf {\bibinfo
  {volume} {23}},\ \bibinfo {pages} {15--41} (\bibinfo {year}
  {2001})}\BibitemShut {NoStop}%
\bibitem [{\citenamefont {Amestoy}\ \emph {et~al.}(2006)\citenamefont
  {Amestoy}, \citenamefont {Guermouche}, \citenamefont {L'Excellent},\ and\
  \citenamefont {Pralet}}]{MUMPS2}%
  \BibitemOpen
  \bibfield  {author} {\bibinfo {author} {\bibfnamefont {P.~R.}\ \bibnamefont
  {Amestoy}}, \bibinfo {author} {\bibfnamefont {A.}~\bibnamefont {Guermouche}},
  \bibinfo {author} {\bibfnamefont {J.-Y.}\ \bibnamefont {L'Excellent}}, \ and\
  \bibinfo {author} {\bibfnamefont {S.}~\bibnamefont {Pralet}},\ }\bibfield
  {title} {\enquote {\bibinfo {title} {Hybrid scheduling for the parallel
  solution of linear systems},}\ }\href@noop {} {\bibfield  {journal} {\bibinfo
   {journal} {Parallel Computing}\ }\textbf {\bibinfo {volume} {32}},\ \bibinfo
  {pages} {136--156} (\bibinfo {year} {2006})}\BibitemShut {NoStop}%
\end{thebibliography}%

\end{document}